\documentclass[fleqn]{aa}
\usepackage{natbib}
\usepackage{morefloats}
\bibpunct{(}{)}{;}{a}{}{,}
\usepackage{graphicx}
\usepackage{amsmath}
\usepackage{amsfonts}
\usepackage{amssymb}

\newlength{\zero}
\newcommand{\z}{\hspace{\zero}}
\newcommand{\zz}{\hspace{\zero}\hspace{\zero}}
\newcommand{\mb}[1]{\mbox{\boldmath $#1$}}
\newcommand{\half}{{1\over2}}

\DeclareMathOperator{\stf}{STF}
\DeclareMathOperator{\imag}{Im}

\begin{document}

\title{General relativistic simulations of passive-magneto-rotational \\ core
  collapse with microphysics}

\author{Pablo Cerd\'a-Dur\'an \inst{1,2}
  \and Jos\'e A. Font \inst{2}
  \and Harald Dimmelmeier \inst{1,3}}

\offprints{Pablo Cerd\'a-Dur\'an, \\ \email{cerda@mpa-garching.mpg.de}}

\institute{Max-Planck-Institut f\"ur Astrophysik,
  Karl-Schwarzschild-Str.~1, 85741 Garching, Germany
  \and
  Departamento de Astronom\'{\i}a y Astrof\'{\i}sica,
  Universidad de Valencia, 46100 Burjassot (Valencia), Spain
  \and
  Department of Physics, Aristotle University of Thessaloniki,
  54124 Thessaloniki, Greece}

\date{Received date / Accepted date}


\abstract{This paper presents results from axisymmetric simulations of
  magneto-rotational stellar core collapse to neutron stars in general
  relativity using the {\it passive field} approximation for the
  magnetic field. These simulations are performed using a new general
  relativistic numerical code specifically designed to study this
  astrophysical scenario. The code is an extension of an existing (and
  thoroughly tested) hydrodynamics code, which has been applied in the
  recent past to study relativistic rotational core collapse. It is
  based on the conformally-flat approximation of Einstein's field
  equations and conservative formulations of the magneto-hydrodynamics
  equations. The code has been recently upgraded to incorporate a
  tabulated, microphysical equation of state and an approximate
  deleptonization scheme. This allows us to perform the most realistic
  simulations of magneto-rotational core collapse to date, which are
  compared with simulations employing a simplified (hybrid) equation
  of state, widely used in the relativistic core collapse
  community. Furthermore, state-of-the-art (unmagnetized) initial
  models from stellar evolution are used. In general, stellar
  evolution models predict weak magnetic fields in the progenitors,
  which justifies our simplification of performing the computations
  under the approach that we call the \emph{passive} field
  approximation for the magnetic field. Our results show that for the
  core collapse models with microphysics the saturation of the
  magnetic field cannot be reached within dynamical time scales by
  winding up the poloidal magnetic field into a toroidal one. We
  estimate the effect of other amplification mechanisms including the
  magneto-rotational instability (MRI) and several types of
  dynamos. We conclude that for progenitors with astrophysically
  expected (i.e.\ weak) magnetic fields, the MRI is the only mechanism
  that could amplify the magnetic field on dynamical time scales. The
  uncertainties about the strength of the magnetic field at which the
  MRI saturates are discussed. All our microphysical models exhibit
  post-bounce convective overturn in regions surrounding the inner
  part of the proto-neutron star. Since this has a potential impact on
  enhancing the MRI, it deserves further investigation with more
  accurate neutrino treatment or alternative microphysical equations
  of state.
  
  }

\authorrunning{Cerd\'a-Dur\'an et al.}
\titlerunning{Relativistic passive-magneto-rotational core collapse}

\maketitle


\section{Introduction}

Understanding the dynamics of the gravitational collapse of the core
of massive stars leading to supernova explosions still remains one of
the primary problems in general relativistic astrophysics, despite the
continuous theoretical efforts during the last few decades. This
problem stands as a distinctive example of a research field where
essential progress has been accomplished through numerical modelling
with increasing levels of complexity in the input physics:
hydrodynamics, gravity, magnetic fields, nuclear physics, equation of
state (EOS), neutrino transport, etc. While studies based upon
Newtonian physics are highly developed nowadays, state-of-the-art
simulations still fail, broadly speaking, to generate successful
supernova explosions under generic conditions (see
e.g.\ \citeauthor{buras} \citeyear{buras}; \citeauthor{kifonidis06}
\citeyear{kifonidis06} for details on the degree of sophistication
achieved in present-day supernova modelling, and
\citeauthor{woosley05} \citeyear{woosley05} and references therein for
a review on the mechanism of core collapse supernovae). The reasons
behind those apparent failures are diverse, all having to do with the
limited knowledge of some of the underlying key issues such as
realistic precollapse stellar models (including rotation, or the
strength and distribution of magnetic fields), the appropriate
EOS, as well as numerical limitations due to the need for Boltzmann
neutrino transport, multi-dimensional hydrodynamics, and relativistic
gravity.

Aside from their assistance to understand the supernova mechanism,
numerical simulations of stellar core collapse are highly motivated
nowadays by the prospects of a direct detection of the gravitational
waves emitted in this scenario. In core collapse events where rotation
plays a role, one of the emission mechanism for gravitational waves is
the hydrodynamic core bounce, which generates a burst signal. The
post-bounce wave signal also exhibits large amplitude oscillations
associated with pulsations in the collapsed core \citep{zwerger_97_a,
rampp_98_a}, neutrino-driven convection behind the supernova shock
\citep{mueller_04_a} and (possibly) rotational dynamical instabilities 
\citep{ott05, ott06a, ott06b}. However, a successful future detection
of gravitational radiation from stellar core collapse faces the
combined problem of the smallness of the signal strength and of the
complexity of the burst signal from bounce. On the other hand, the
energy released in gravitational waves is so small that its
backreaction to the collapse dynamics is negligible, which can
significantly simplify the numerical simulation of this scenario. To
pave the road for a successful detection through waveform templates
for data analysis, such simulations are essential.

At birth neutron stars have intense magnetic fields
($ \sim 10^{12} \mbox{\,--\,} 10^{13} \mathrm{\ G} $) or in extreme
cases even larger ones
($ \sim 10^{14} \mbox{\,--\,} 10^{15} \mathrm{\ G} $),
as inferred from studies of anomalous X-ray pulsars and soft gamma-ray
repeaters \citep{kouveliotou98}. For magnetars, the magnetic field can
be so strong as to alter the internal structure of the neutron star
\citep{bocquet95}. The emergence of such strong magnetic fields in
neutron stars from the initial field configuration in the pre-collapse
stellar cores is an active and important field of research. Similarly,
the rotation state of the nascent proto-neutron star (PNS) is
determined by the amount and distribution of angular momentum in the
core of the progenitor, which is still rather unconstrained, being only
currently incorporated into stellar evolution codes \citep{heger05}.
Observations of surface velocities imply that a large fraction of
progenitor cores is rapidly rotating. The presence of intense magnetic
fields, on the other hand, may also affect rotation in the core, as it
can be spun down in the red giant phase by magnetic torques via dynamo
action which couples to the outer layers of the star \citep{meier76,
  spruit98, spruit02, heger05}. The latest numerical calculations of
stellar evolution thus predict low pre-collapse core rotation rates,
which are in agreement with observed periods of young neutron stars in
the range of $ \sim 10 \mbox{\,--\,} 15 \mathrm{\ ms} $. Nevertheless,
a recent estimate by \citet{woosley06} indicates that $ \sim 1\% $ of
all stars with $ M \ge 10 M_\odot $ could still have rapidly rotating
cores, which could also be relevant for the collapsar-type gamma-ray
burst scenario.

The presence of intense magnetic fields in a PNS makes
magneto-rotational core collapse simulations mandatory. The weakest
point of all existing simulations to date is the fact that both the
strength and distribution of the initial magnetic field in the core
are basically unknown. If the magnetic field is initially weak, there
exist several mechanisms which may amplify it to values which can have
an impact on the dynamics, among them differential rotation ($ \Omega
$-dynamo\footnote{Note that the ``$ \Omega $-dynamo'' is also referred
  to in the literature as ``wind-up'' or ``field-wrapping''. We follow
  in this paper the notation used by
  \citet{obergaulinger_06_b,obergaulinger_06_a}}, the magneto-rotational
instability (MRI hereafter), as well as dynamo mechanisms related to
convection or turbulence. The first of these mechanisms transforms
rotational energy into magnetic energy, winding up any seed poloidal
field into a toroidal field. The MRI leads to a sustained turbulent
dynamo which is able to transport angular momentum outwards, although
it remains unclear how large the actual amplification by this process
can be (see below). The latter mechanisms, which are generically
called $\alpha\mbox{-}\Omega$-dynamo and will be discussed below,
include a number of processes which can also produce an amplification
of the magnetic field.

Magneto-rotational core collapse simulations were first performed as
early as in the 1970s \citep{leblanc70, bisnovatyi76, meier76,
mueller79, ohnishi83, symbalisty84}, in which magneto-rotational core
collapse was already proposed as a plausible supernova explosion
mechanism. In recent years, an increasing number of authors have
performed axisymmetric magneto-hydrodynamic (MHD) simulations of
stellar core collapse (within the so-called ideal MHD limit) employing
a Newtonian treatment of MHD and gravity, and either a simplified
equation of state \citep{yamada04, ardeljan05, sawai05} or a
microphysical description of matter \citep{kotake04_a, kotake04_b,
kotake05}. The main implications of the presence of strong magnetic
fields in the collapse are the redistribution of the angular momentum
and the appearance of jet-like explosions. Specific magneto-rotational
effects on the gravitational wave signature were first studied in
detail by \citet{kotake04_a} and \citet{yamada04}, who found
differences with purely hydrodynamic models only for very strong
initial fields ($ \geq 10^{12} \mathrm{\ G} $). The most exhaustive
parameter study of magneto-rotational core collapse to date has been
carried out very recently by \citet{obergaulinger_06_b,
  obergaulinger_06_a}. Their axisymmetric simulations employed
rotating polytropes, Newtonian hydrodynamics and gravity
(approximating general relativistic effects via an effective
relativistic gravitational potential in their latter work), and ad-hoc
initial poloidal magnetic field distributions. These authors found
that for weak initial fields ($ \leq 10^{11} \mathrm{\ G} $, which is
the astrophysically most motivated case) there are no differences
compared to purely hydrodynamic simulations, neither in the collapse
dynamics nor in the resulting gravitational wave signal. However,
strong initial fields ($ \geq 10^{12} \mathrm{\ G} $) manage to slow
down the core efficiently (leading even to retrograde rotation in the
PNS) which causes qualitatively different dynamics and gravitational
wave signals. For the most strongly magnetized models
\citet{obergaulinger_06_a} found highly bipolar, jet-like outflows.

Nowadays, there exists sophisticated numerical technology to
perform general relativistic hydrodynamics simulations \citep[see
e.g.][]{font_03_a}. In recent years this technology has been extended
to general relativistic magneto-hydrodynamics (GRMHD)
\citep{Koide99,DeVilliers03,delzanna03,Gammie03,Duez05,anton06}. General
relativistic simulations involve the challenging computational task of
solving Einstein's field equations coupled to the fluid (and
magneto-fluid) evolution. The first general relativistic axisymmetric
simulations of rotational stellar core collapse to neutron stars were
performed by \citet{dimmelmeier_01_a, dimmelmeier_02_a,
dimmelmeier_02_b}. These simulations employed simplified models to
describe the thermodynamics of the process, in the form of a
polytropic EOS modified such that it accounts both for the stiffening
of the matter above nuclear density as well as thermal heating by the
passing shock front \citep[the so-called hybrid EOS;
see][]{janka_93_a}. The inclusion of relativistic effects within the
so-called CFC approximation results primarily in a stronger gravitational
pull during the contraction of the core. Thus, higher densities than
in Newtonian models are reached during bounce, and the nascent PNS is
more compact. Relativistic simulations with improved dynamics and
gravitational waveforms were reported by \citet{cerda05}, who used the
CFC+ framework, which includes second post-Newtonian corrections to
CFC. Comparisons of the CFC approach with fully general relativistic
simulations (employing also stable reformulations of the Einstein
equations in $ 3 + 1 $ form) have been reported by
\citet{shibata_04_a}, \citet{ott06a}, and \citet{ott06b} in the
context of axisymmetric core collapse simulations. As in the case of
CFC+, the differences found in both the collapse dynamics and the
gravitational waveforms are minute, which highlights the suitability
of CFC for performing accurate simulations of such scenarios without
the need for solving the full system of Einstein's equations. Owing to
the excellent approximation of full general relativity offered by CFC
in the context of stellar core collapse, extensions to improve the
microphysics through the incorporation of a tabulated non-zero
temperature EOS and a simplified neutrino treatment have been
recently reported by \citet{ott06a} and \citet{dimmelmeier_07_a}.
These simulations allow a direct comparison to the models presented in
\citet{dimmelmeier_02_b}, \citet{cerda05}, and \citet{shibata_04_a},
which use the same parameterization of rotation but a simple hybrid
EOS. This comparison shows that with a microphysical treatment the
influence of rotation on the collapse dynamics and waveforms is
significantly reduced. In particular, the most important result of
\cite{dimmelmeier_07_a} is the suppression of core collapse with multiple
centrifugal bounces and its associated Type II gravitational waveforms
\citep[see][]{dimmelmeier_02_b}.

On the other hand, to further improve the realism of core collapse
simulations in general relativity, the incorporation of magnetic
fields in numerical codes via solving the MHD equations is also
currently being undertaken \citep{shibata_06_a, nfnr}. The work of
\citet{shibata_06_a} is focused on the collapse of initially strongly
magnetized cores ($ \sim 10^{12} \mbox{\,--\,} 10^{13} \mathrm{\ G} $).
Although these values are probably astrophysically not relevant
\citep[as the stellar evolution models of][predict a poloidal field
strength of $ \sim 10^6\mathrm{\ G} $ in the progenitor]{heger05},
they enable them  to resolve the scales needed for the MRI to develop,
since the MRI length scale grows with the magnetic field. The results
of \citet{shibata_06_a} show that the magnetic field is mainly
amplified by the wind-up of the magnetic field lines by differential
rotation. Consequently, the magnetic field is accumulated around the
inner region of the PNS, and a MHD outflow forms along the rotation
axis removing angular momentum from the PNS. A different approach is
followed by \citet{nfnr}. Their progenitors are chosen to be weakly
magnetized ($ \le 10^{10} \mathrm{\ G} $) which is in much better
agreement with predictions from stellar evolution. Under these
conditions the so-called ``passive'' magnetic field approximation (see
Sect.~\ref{sec:passive} below) is appropriate. In addition, the
numerical code used in that work, which utilizes spherical
coordinates, is more suitable for core collapse simulations than codes
based on Cartesian/cylindrical coordinates, as used e.g.\ by
\citet{shibata_06_a}.

In this paper we continue the program initiated in \citet{nfnr} to
build a numerical code which includes all relevant ingredients to
study relativistic magneto-rotational stellar core collapse. To this
aim we present here the first relativistic simulations of
magneto-rotational core collapse which take into account the effects
of a microphysical EOS and a simplified neutrino treatment. Those
effects have been incorporated in the code following the approach
recently presented by \citet{ott06a} and \citet{dimmelmeier_07_a}.
As in \citet{nfnr} we employ the passive magnetic field approximation
in the treatment of the magnetic field.

The paper is organized as follows: Sect.~\ref{theory} presents a
brief overview of the theoretical framework we use to perform
relativistic simulations of core collapse. Sect.~\ref{sec:mag_ini_models}
describes how the magnetized initial models for core collapse are built
and presents our sample of models. In Sect.~\ref{eos} we discuss
aspects related to incorporating microphysics in the core collapse
models and their implementation in the numerical code. A brief outline
of our numerical approach is discussed in Sect.~\ref{numerics}. The
evolution of the core collapse initial models is discussed in
Sect.~\ref{results}. The main paper closes with a summary in
Sect.~\ref{sec:conclusions}. Relevant tests of the code are analyzed
in Appendix~\ref{app:tests}, while Appendix~\ref{app:odynamo} provides
an estimate for the growth rate of the $ \Omega $-dynamo.

Throughout the paper we use a spacelike signature $ (-, +, +, +) $ and
units in which $ c = G = 1 $. Greek indices run from 0 to 3, Latin
indices from 1 to 3, and we adopt the standard Einstein summation
convention.


\section{Theoretical framework}
\label{theory}

We adopt the $ 3 + 1 $ formalism of general relativity
\citep{lichnerowicz44} to foliate the spacetime into spacelike
hypersurfaces. In this approach the line element reads
\begin{equation}
  \mathrm{d}s^2 = - \alpha^2 \, \mathrm{d}t^2 +
  \gamma_{ij} (\mathrm{d}x^i + \beta^i \,\mathrm{d}t)
  (\mathrm{d}x^j + \beta^j \, \mathrm{d}t),
\end{equation}
where $ \alpha $ is the lapse function, $ \beta^i $ is the shift
vector, and $ \gamma_{ij} $ is the spatial three-metric induced in
each hypersurface. Using the projection operator $ \perp^\mu_\nu $ and
the unit four-vector $ n^\mu $ normal to each hypersurface, it is
possible to build the quantities
\begin{eqnarray}
  E & = & n^{\mu} n^{\nu} T_{\mu\nu} = \alpha^2 T^{00},
  \label{eq:tmn_projection_1}
  \\
  S_i & = & - \perp^{\mu}_{i} n^{\nu} T_{\mu\nu} =
  - \frac{1}{\alpha} (T_{0i} - T_{ij} \beta^j),
  \label{eq:tmn_projection_2}
  \\
  S_{ij} & = & \perp^{\mu}_i \perp^{\nu}_j T_{\mu\nu} = T_{ij},
  \label{eq:tmn_projection_3}
\end{eqnarray}%
which represent the total energy, the momenta, and the spatial
components of the stress-energy tensor, respectively.

To solve the gravitational field equations we choose the ADM gauge in
which the three-metric can be decomposed as
$ \gamma_{ij} = \phi^4 \hat{\gamma}_{ij} + h^\mathrm{TT}_{ij}$, where
$ \phi $ is the conformal factor, $\hat{\gamma}_{ij}$ is the flat
three-metric, and $ h^\mathrm{TT}_{ij} $ is the transverse and
traceless part of the three-metric. Note that this gauge choice
implies the maximal slicing condition in which the trace $ K $ of the
extrinsic curvature tensor $ K_{ij} $ vanishes.


\subsection{The CFC approximation}

In our work Einstein's field equations are formulated and solved using
the conformally flat condition (CFC hereafter), introduced by
\citet{isenberg_78_a} and first used in a dynamical context by
\citet{wilson_96_a}. In this approximation, the three-metric in the
ADM gauge is assumed to be conformally flat,
$ \gamma_{ij} = \phi^4 \hat\gamma_{ij} $. Note that the same
aproximation can be realized for other gauge choices such as the
quasi-isotropic gauge or the Dirac gauge, both supplemented by the
maximal slicing condition. Under the CFC assumption the gravitational
field equations can be written as a system of five nonlinear
elliptic equations,
\begin{eqnarray}
  \hat{\Delta} \phi & = & - 2 \pi \phi^5
  \left( E + \frac{K_{ij}K^{ij}}{16 \pi} \right),
  \label{eq:cfc1}
  \\
  \hat{\Delta} (\alpha \phi) & = & 2 \pi \alpha \phi^5
  \left( E + 2 S + \frac{7 K_{ij}K^{ij}}{16 \pi} \right),
  \label{eq:cfc2}
  \\
  \hat{\Delta} \beta^i & = & 16 \pi \alpha \phi^4 S^i +
  2 \phi^{10} K^{ij} \hat{\nabla}_j
  \left(\! \frac{\alpha}{\phi^6} \!\right) -
  \frac{1}{3} \hat{\nabla}^i \hat{\nabla}_k \beta^k,
  \label{eq:cfc3}
\end{eqnarray}
where $ \hat{\Delta} $ and $ \hat{\nabla} $ are the Laplace and nabla
operators associated with the flat three-metric, and
$ S = \gamma^{ij} S_{ij} $.


\subsection{General relativistic magnetohydrodynamics}

The energy-momentum tensor of a magnetized perfect fluid can be
written as the sum of the fluid part and the electromagnetic field
part. In the so-called ideal MHD limit (where the fluid is a perfect
conductor of infinite conductivity), the latter can be expressed
solely in terms of the magnetic field $ b^\mu $ measured by a comoving
observer. In this case the total energy-momentum tensor is given by
\begin{equation}
  T^{\mu \nu} = (\rho h + b^2) \, u^\mu u^\nu +
  \left( P + \frac{b^2}{2} \right) g^{\mu \nu} - b^\mu b^\nu,
  \label{eq:tmunu_grmhd}
\end{equation}
where $ \rho $ is the rest-mass density, $ h = 1 + \epsilon + P / \rho $
the relativistic enthalpy, $ \epsilon $ the specific internal
energy, $ P $ the pressure, and $ u^\mu $ the four-velocity of the
fluid, while $ b^2 = b^\mu b_\mu $. We define the magnetic pressure
$ P_\mathrm{mag} = b^2 / 2 $ and the specific magnetic energy
$ \epsilon_\mathrm{mag} = b^2 / (2 \rho) $, whose effect on the
dynamics is similar to that played by the pressure and specific
internal energy of the fluid, respectively. In the ideal MHD limit,
the electric field measured by a comoving observer vanishes, and
Maxwell's equations simplify. Under this assumption the electric field
four-vector $ E^\mu $ can be expressed in terms of the magnetic field
four-vector $ B^\mu $, and only equations for $ B^i $ are
needed. For an Eulerian observer, $ U^\mu = n^\mu $, the temporal
component of the electric field vanishes,
$ E^\mu = (0, - \varepsilon_{ijk} v^j B^k) $. In this case Maxwell's
equations reduce to the divergence-free condition and the induction
equation for the magnetic field,
\begin{equation}
  \hat{\nabla}_i B^{*\,i} = 0,
  \qquad
  \frac{\partial B^{*\,i}}{\partial t} =
  \hat{\nabla}_j (v^{*\,i} B^{*\,j} - v^{*\,j} B^{*\,i}),
\end{equation}
where $ B^{*\,i} = \sqrt{\bar{\gamma}} B^i $ and
$ v^{*\,i} = \alpha v^i - \beta^i $, with $ v^i $ being the fluid's
three-velocity as measured by the Eulerian observer. The ratio of the
determinants of the three-metric and the flat three-metric is given by
$ \bar{\gamma} = \gamma / \hat{\gamma} $. In the Newtonian limit
$ v^{*\,i} \to v^i $ and $ B^{*\,i} \to B^i $, and the Newtonian induction
equation and divergence constraint are recovered.

The evolution of a magnetized fluid is determined by the conservation
law of the energy-momentum, $ \nabla_\mu T^{\mu \nu} = 0 $, and by the
continuity equation, $ \nabla_\mu J^\mu = 0 $, for the rest-mass
current $ J^\mu = \rho u^\mu $. Following the procedure laid out in
\citet{anton06}, in order to cast the GRMHD equations as a hyperbolic system
of conservation laws well adapted to numerical work, the conserved
quantities are chosen in a way similar to the purely hydrodynamic case
presented by \citet{banyuls_97_a}:
\begin{eqnarray}
  D & = & \rho W,
  \\
  S_i & = & (\rho h + b^2) W^2 v_i - \alpha b_i b^0,
  \\
  \tau & = & (\rho h + b^2) W^2 - \left( P + \frac{b^2}{2} \right) -
  \alpha^2 (b^0)^2 - D,
\end{eqnarray}%
where $ W = \alpha u^0 $ is the Lorentz factor. With this choice the
system of conservation equations for the fluid and the induction
equation for the magnetic field can be cast as a first-order,
flux-conservative, hyperbolic system,
\begin{equation}
  \frac{\partial \sqrt{\gamma} \mb{U}}{\partial t} +
  \frac{\partial \sqrt{- g} \mb{F}^i}{\partial x^i} =
  \sqrt{- g} \mb{S},
  \label{eq:hydro_conservation_equation}
\end{equation}
with the state vector, flux vector, and source vector given by
\begin{eqnarray}
  \mb{U} & = & [D, S_j, \tau, B^k],
  \label{eq:state_vector}
  \\
  \mb{F}^i & = & \left[
  D \hat{v}^i, S_j \hat{v}^i + \delta^i_j \left( P + \frac{b^2}{2} \right) -
  \frac{b_j B^i}{W}, \right.
  \nonumber
  \\
  & & \left. \:\: \tau \hat{v}^i + \left( P + \frac{b^2}{2} \right) v^i -
  \alpha \frac{b^0 B^i}{W}, \hat{v}^i B^k - \hat{v}^k B^i \right],
  \label{eq:flux_vector}
  \\
  \mb{S} & = & \left[ 0, \frac{1}{2} T^{\mu \nu}
  \frac{\partial g_{\mu \nu}}{\partial x^j},
  \alpha \! \left( \! T^{\mu 0} \frac{\partial \ln \alpha}{\partial x^\mu} -
  T^{\mu \nu} {\it \Gamma}^0_{\mu \nu} \! \right), 0^k \! \right],
  \label{eq:source_vector}
\end{eqnarray}%
where $ \delta^i_j $ is the Kronecker delta and
$ \Gamma^\mu_{\mu \lambda}$ are the Christoffel symbols associated with
the four-metric. We note that  the above definitions contain
components of the magnetic field measured by both a comoving observer
and an Eulerian observer. The two are related by
\begin{equation}
  b^0 = \frac{W B^i v_i}{\alpha},
  \qquad
  b^i = \frac{B^i + \alpha b^0 u^i}{W}.
\end{equation}

The hyperbolic structure of Eq.~(\ref{eq:hydro_conservation_equation})
and the associated spectral decomposition (into eigenvalues and
eigenvectors) of the flux-vector Jacobians is given in
\citet{anton06}. This information is needed for numerically solving
the system of equations using the class of high-resolution
shock-capturing schemes that we have implemented in our code.


\subsection{The passive field approximation}
\label{sec:passive}

In the collapse of \emph{weakly magnetized} stellar cores, it is a
fair approximation to assume that the magnetic field entering in the
energy-momentum tensor of Eq.~(\ref{eq:tmunu_grmhd}) is negligible
when compared with the fluid part, i.e.\ $ P_\mathrm{mag} \ll P $,
$ \epsilon_\mathrm{mag} \ll \epsilon $, and that the components of the
anisotropic term of $ T^{\mu \nu} $ satisfy
$ b^{\mu}b^{\nu} \ll \rho h u^{\mu} u^{\nu} + P g^{\mu \nu} $. With
these simplifications the evolution of the magnetic field, governed by
the induction equation, does not affect the dynamics of the fluid,
which is governed solely by the hydrodynamics equations. However, the
magnetic field evolution does depend on the fluid evolution, due to
the presence of the velocity components in the induction equation.
This ``test magnetic field'' (or passive field) approximation is
employed in the core collapse simulations reported in this work.

Within this approach the seven eigenvalues of the GRMHD Riemann
problem (entropy, Alfv\'en, and fast and slow magnetosonic waves)
reduce to three \citep{nfnr},
\begin{eqnarray}
  \lambda^{i}_{0 \, \mathrm{hydro}} & = &
  \lambda^i_\mathrm{e} = \lambda^i_{\mathrm{A} \, \pm} =
  \lambda^i_{\mathrm{s} \, \pm},
  \\
  \lambda^i_{\pm \, \mathrm{hydro}} & = &
  \lambda^i_{\mathrm{f} \, \pm},
\end{eqnarray}
where $ \lambda^{i}_{0 \, \mathrm{hydro}} $ and
$ \lambda^i_{\pm \, \mathrm{hydro}} $ are the eigenvalues of the
Jacobian matrices of the purely hydrodynamics equations, as reported
by \citet{banyuls_97_a}.

This approximation has several interesting properties. First, if we
perform a simulation for a given initial magnetic field, we can
compute the result for a simulation with the same initial magnetic
field scaled by some factor. To do this it is sufficient to increase
or reduce the strength of the magnetic field at any given time during
the simulation by the same factor. The second property is what we call
the ``composition rule''. If we perform two simulations with the same
hydrodynamics but different initial magnetic fields,
$ \mb{B}^{*\,0}_1 $ and $ \mb{B}^{*\,0}_2 $, any linear combination
$ \mb{B}^* (t)= a \, \mb{B}^*_1 (t) + b \, \mb{B}^*_2 (t) $ of the
magnetic field at any time, with $ a $ and $ b $ being constants, will
be the solution for the evolution of a model whose initial magnetic
field is the same linear combination, i.e.\
$ \mb{B}^{*\,0}= a \, \mb{B}^{*\,0}_1 + b \, \mb{B}^{*\,0}_2 $. Hence,
we can make use of these properties to cover a wide range of magnetic
field strengths and structures by performing just a few simulations,
and then constructing additional ones by means of the ``composition
rule''. Needless to say, these two properties are valid only if the
magnetic field resulting from the scaling or composition satisfies
itself the passive field approximation at all times.


\subsection{Gravitational waves}

The Newtonian standard quadrupole formula has been extensively used in
numerical simulations of astrophysical systems to compute the
gravitational radiation and waveforms without having to consider the
full evolution of the spacetime and solving Einstein's equations.
This formula computes the radiative part of the spatial metric as
\begin{equation}
  h^\mathrm{quad}_{ij} =
  P^{\mathrm{TT} kl}_{ij} \frac{2}{R} \ddot{Q}_{ij},
  \label{eq:quad_formula}
\end{equation}
where $ P^{\mathrm{TT} kl}_{ij} $ is the transverse traceless
projector operator \citep{thorne80}, $ R $ is the distance to the
source, $ Q_{ij} $ is the mass quadrupole moment, and a dot denotes a
time derivative. In spite of its simplicity, the particular form in
which Eq.~(\ref{eq:quad_formula}) is expressed leads to numerical
difficulties due to the presence of second time derivatives. A way to
circumvent this problem is to eliminate all time derivatives using the
equations of motion. Following \citet{finn_89_a} and
\citet{blanchet_90_a} one can arrive to an expression for
$ \ddot{Q}_{kl} $ with no explicit appearance of time derivatives.
This is the so-called stress formula,
\begin{equation}
  \ddot{Q}_{ij} \approx \stf \left\{ 2 \int \mathrm{d}^3 \mb{x} \,
  \sqrt{\hat{\gamma}} D^* \left( \hat{\gamma}_{ik} \hat{\gamma}_{jl}
  \, v^k v^l +  x^k \hat{\gamma}_{ki} \, \hat{\nabla}_j U \right) \right\},
  \label{eq:stress_formula}
\end{equation}
where $ \stf $ means the symmetric and traceless part, and 
$ D^* = \sqrt{\bar{\gamma}} D $. This formula has proved to be
numerically much more accurate than the original formula and we use it
in this paper to extract gravitational waveforms.

In the case of a magnetized fluid in the ideal MHD case, the
gravitational radiation is also affected by the energetic content of
the magnetic field. \citet{kotake04_b} have derived an extension of
the quadrupole formula for such a case. In a similar way, it is
possible to calculate the corresponding stress formula
\citep{obergaulinger_06_a}, which reads
\begin{eqnarray}
  \ddot{Q}_{ij} & \approx & \stf
  \biggl\{ 2 \int \mathrm{d}^3 \mb{x} \, \sqrt{\hat{\gamma}} \biggl[ D^*
  \left( \hat{\gamma}_{ik} \hat{\gamma}_{jl} \, v^k v^l +
  x^k \hat{\gamma}_{ki} \, \hat{\nabla}_j U \right)
  \nonumber
  \\
  & & \qquad \qquad \qquad \qquad \,\;
  - \hat{\gamma}_{ik} \hat{\gamma}_{jl} \, b^k b^l \biggr] \biggr\}.
  \label{eq:mag_stress_formula}
\end{eqnarray}
Note that in the limit of weak magnetic fields the original stress
formula is recovered. We use this formula in the magnetized core
collapse simulations to calculate the contribution of the magnetic
field to the waveforms in the passive field approximation.


\section{Initial data}
\label{sec:mag_ini_models}

The structure and strength of the magnetic field in the stellar core
collapse progenitors, needed as initial conditions of our numerical
simulations, is still an open question in
astrophysics. State-of-the-art models from stellar evolution including
a description for the influence of the magnetic field
\citep{heger05}, predict that the distribution of the magnetic field in
the iron core has probably a dominant toroidal component, with a strength of
about $ 10^9 \mbox{\,--\,} 10^{10} \mathrm{\ G} $, and a poloidal
component of only about $ 10^5 \mbox{\,--\,} 10^6 \mathrm{\ G} $.
For such weak fields ($ P_\mathrm{mag} \ll P $),
the passive field approximation adopted here is likely to be
sufficient to describe the initial models considered in this work. A
second consideration is whether or not the initial model should be an
equilibrium model. In general, if one tries to construct a stationary
model without meridional currents and assuming an isentropic flow, the
only possible magnetic field configuration is poloidal
\citep[see][]{bekenstein79}. Stationary models of magnetized stars
have been computed under these assumptions by \citet{bocquet95}. In
the general (but still isentropic) case in which meridional
circulation is allowed, a toroidal component of the magnetic field may
exist, but the method to calculate stationary models is far more
complicated \citep{gourgulhon93, ioka03, ioka04}. When one considers
ideal MHD, also purely toroidal magnetic fields exist which maintain
the circularity condition \citep{oron02}, and therefore it is possible
to generate stationary models without meridional components. Finally,
in the case that magnetic pressure does not exceed the hydrostatic
pressure, \citet{oron02} has shown that stationary models with mixed
toroidal and poloidal component approximately accomplish the
circularity condition.

Therefore, it makes sense to construct initial models for magnetized
stellar cores by simply adding an ad-hoc magnetic field to a purely
hydrodynamic equilibrium configuration. If the condition
$ \vec{B}^* \cdot \vec{\hat{\nabla}} \Omega^* = 0 $ is satisfied, where
$ \Omega^* = v^{*\,\varphi} / (r \sin{\theta}) $ is the angular
velocity of the fluid,  then the initial magnetic field
does not evolve in time either. Note that in this work we use as initial
models both equilibrium and non-equilibrium configurations for the
magnetic field, as specified in Table~\ref{tab:MCC_models}.


\subsection{Magnetic field configurations}

Since the numerical scheme we use to evolve the MHD equations only
preserves the value of $ \vec{\hat{\nabla}} \cdot \vec{B}^* $ but does
not impose the divergence constraint of the magnetic field itself, it
is necessary to build initial configurations that also satisfy this
condition. To do this we calculate the initial magnetic field from a
vector potential $ \vec{A}^* $, such that
$ \vec{B}^* = \vec{\hat{\nabla}} \times \vec{A}^* $, which is
discretized as explained in \citet{nfnr}.

For our code tests and core collapse simulations we use three possible
magnetic field configurations as initial conditions (or any possible
combination of them):

\noindent
-- the homogeneous ``starred'' magnetic field, in which $ \vec{B}^* $
is constant and parallel to the symmetry axis,

\noindent
-- the poloidal magnetic field generated by a circular current loop
of radius $ r_\mathrm{mag} $ \citep{jackson62}, that can be calculated
from the only non-vanishing component of the vector potential
$ A^*_\varphi $ as
\begin{equation}
  A^*_\varphi = \frac{r^2_\mathrm{mag} B^*_0}{2}
  \sum_{n=0}^{\infty} \frac{(-1)^n (2n-1)!!}{2^n (n+1)!}
  \frac{r_<^{2n+1}}{r_>^{2n+2}} P^1_{2n+1} (\cos{\theta}),
  \label{eq:circ_loop}
\end{equation}
where $ r_< = \min(r, r_\mathrm{mag}) $,
$ r_> = \max(r, r_\mathrm{mag}) $, and $ B^*_0 $ is the
magnetic field at the center, and

\noindent
-- a toroidal magnetic field of the form
\begin{equation}
  B^{*\,\varphi} =
  B^*_0 \frac{r_\mathrm{mag}^2}{r_\mathrm{mag}^2 - \varpi^2},
  \label{eq:pure_tor}
\end{equation}
whose maximum value is reached at $ \varpi = r_\mathrm{mag} $, where
$ B^*_0 $ is the initial central magnetic field and $ \varpi $ is the
distance to the axis.

\begin{figure*}[t!]
  \centering
  \resizebox{0.49\textwidth}{!}{\includegraphics*{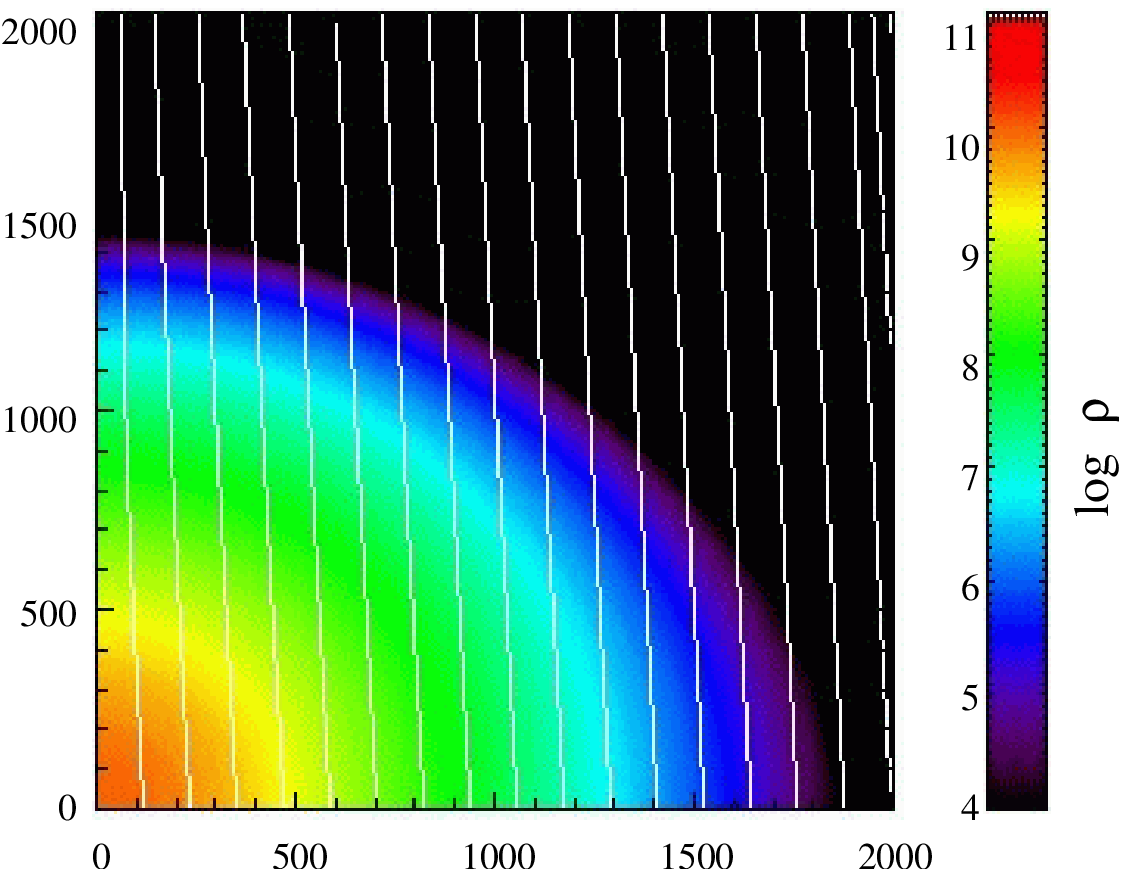}}
  \resizebox{0.49\textwidth}{!}{\includegraphics*{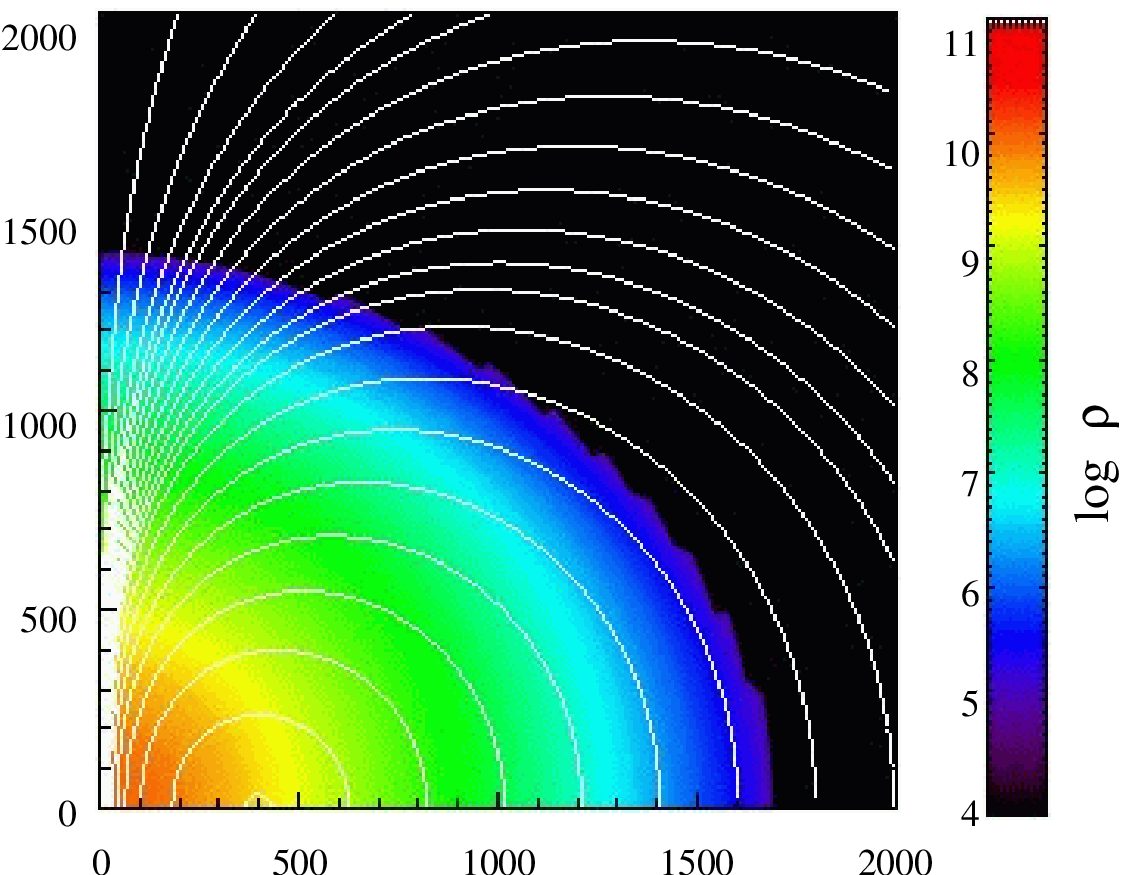}}
  \caption{Logarithm of the rest mass density ($ \log{\rho} $, color
    coded) and magnetic field lines ($ B^i $, white curves) in a
    homogeneous $ B^{*\,i} $ field configuration (left panel) and a
    magnetic field generated by a circular current loop (right
    panel), for a typical rotating iron core (model A1B3 in
    Table~\ref{tab:CC_models}) used for the collapse simulations. The
    magnetic field lines have been calculated as isocontours of the
    vector potential. The axis scale is in km and the density in
    cgs units.}
  \label{fig:homo_field}
\end{figure*}

Note that in all three cases, we employ the ``starred'' magnetic
field, since the divergence constraint is valid for this quantity when
computed with respect to the flat divergence operator. In this way we
can extend any analytic prescription for the magnetic field given in
flat spacetime in an easy way. Also note that in the presence of strong
gravitational fields the magnetic field $ \vec{B} $ is deformed with
respect to $ \vec{B}^* $ due to the curvature of the spacetime,
although the divergence constraint is automatically fulfilled.
Fig.~\ref{fig:homo_field} shows examples of the magnetic field
structure for the poloidal configurations of the initial magnetic
field.

The initial magnetic field configuration is denoted in the names of
the models in our sample by adding a label to the purely
hydrodynamic model. For the latter we follow the notation of
\citet{dimmelmeier_02_a} and \citet{ott06a}. These models are listed in
Table~\ref{tab:CC_models}. The label for the magnetic field is
constructed following the notation of \citet{obergaulinger_06_a}. We
add the suffix D3M0 to denote those models with purely poloidal
magnetic field generated by a circular loop and
$ r_\mathrm{mag} = 400 \mathrm{\ km} $ (M0 denotes the passive field
approximation) and we use the suffix T3M0 for models with purely
toroidal magnetic field and $ r_\mathrm{mag} = 400 \mathrm{\ km} $.
We have also built the DT3M0 model, whose magnetic field distribution
is a combination of D3M0 and T3M0 with equal magnetic field strengths
at the center. This model allows us to check the validity of the
``composition rule'' (see Sect.~\ref{sec:passive}).


\subsection{Hydrodynamic configurations}


\subsubsection{Polytropes in rotational equilibrium}

For the simulation with a simplified description of matter using the
hybrid EOS (see Sect.~\ref{hybrid_eos}), we construct
$ \gamma_\mathrm{ini} = 4/3 $ polytropes in rotational equilibrium
which we obtain by using the relativistic generalization of Hachisu's
self-consistent field method by \citet{komatsu_89_a}\footnote{The
  adiabatic index should not be confused with the determinant of the
  spacetime three-metric, although we use the same symbol $ \gamma $
  (following usual practice).}. Their rotation law for the specific
angular momentum $ j $ is given by
\begin{equation}
  j = A^2 (\Omega_\mathrm{c} - \Omega),
  \label{rotation_law}
\end{equation}
where $ A $ parametrizes the degree of differential rotation (stronger
differentiality with decreasing $ A $) and $ \Omega_\mathrm{c} $ is
the value of the angular velocity $ \Omega $ at the center. In the
Newtonian limit, this reduces to
\begin{equation}
  \Omega = \frac{A^2 \Omega_\mathrm{c}}{A^2 + \varpi^2}.
  \label{newtonian_rotation_law}
\end{equation}
The parameters of the selected models, which are chosen to be
identical to some of the models considered by
\citet{dimmelmeier_02_a}, are described in Table~\ref{tab:CC_models}.
In addition, as we aim at comparing our results with the recent
numerical simulations performed by \citet{obergaulinger_06_a} in
Newtonian gravity, a subset of our models (those with purely poloidal
magnetic field) have been selected as general relativistic
counterparts of their models. In Table~\ref{tab:CC_models} we also
give the values for the gravitational mass $ M_\mathrm{g} $ (which is
identical to the ADM mass $ M_\mathrm{ADM} $) and for the initial
rotation rate $ \beta = E_\mathrm{rot} / |E_\mathrm{b}| $. In the
definition of $ \beta $ we use the following expressions for the
rotational kinetic energy $ E_\mathrm{rot} $, the gravitational
binding energy $ E_\mathrm{b} $, and the magnetic energy
$ E_\mathrm{mag} $:
\begin{eqnarray}
  E_\mathrm{rot} & = & \frac{1}{2} \int \mathrm{d}^3 \mb{x} \,
  \sqrt{\gamma} \, \alpha \hat{v}^\varphi S_\varphi,
  \\
  E_\mathrm{b} & = & M_\mathrm{g} - M_\mathrm{p} -
  E_\mathrm{rot} - E_\mathrm{mag},
  \\
  E_\mathrm{mag} & = & \frac{1}{2} \int \mathrm{d}^3 \mb{x} \,
  \sqrt{\gamma} \, W b^2,
\end{eqnarray}%
where $ M_\mathrm{p} $ is the proper mass.

\begin{table}[t!]
  \centering
  \caption{Purely hydrodynamic initial models used in the magnetized
    core collapse simulations.}
  \begin{tabular}{lccccc}
    \hline \hline
    Model
    & $ \rho_\mathrm{c} $
    & \multicolumn{1}{c}{$ A $}
    & $ \beta $
    & $ M_\mathrm{g} $
    & $ \gamma_1 $
    \\
    & [$ 10^{10} \mathrm{\, g\, cm}^{-3} $]
    & \multicolumn{1}{c}{[$ 10^3 \mathrm{\ km} $]}
    & [\%]
    & [$ M_\odot $] &
    \\
    \hline
    A1B3G3  & 1.00 &  50.0 & 0.90 & 1.46 & 1.300 \\
    A1B3G5  & 1.00 &  50.0 & 0.90 & 1.46 & 1.280 \\
    A3B3G5  & 1.00 & \z0.5 & 0.90 & 1.46 & 1.280 \\
    A2B4G1  & 1.00 & \z1.0 & 1.80 & 1.50 & 1.325 \\
    A4B5G5  & 1.00 & \z0.5 & 4.00 & 1.61 & 1.280 \\
    s20A1B1 & 0.88 &  50.0 & 0.25 & 1.58 & ---   \\
    s20A1B5 & 0.88 &  50.0 & 4.00 & 1.58 & ---   \\
    s20A2B2 & 0.88 & \z1.0 & 0.50 & 1.58 & ---   \\
    s20A2B4 & 0.88 & \z1.0 & 1.80 & 1.58 & ---   \\
    s20A3B3 & 0.88 & \z0.5 & 0.90 & 1.58 & ---   \\
    E20A    & 0.42 & ---   & 0.37 & 2.00 & ---   \\
    \hline \hline
  \end{tabular}
  \label{tab:CC_models}
\end{table}

For these simplified initial models the gravitational collapse is
initiated by slightly decreasing the adiabatic index from its initial
value to $ \gamma_1 < \gamma_\mathrm{ini} = 4 / 3 $, which results in a
loss of pressure support. If no pressure reduction were imposed,
the purely hydrodynamic initial models would remain
stationary. However, even in that case the associated initial magnetic
field may not remain stationary (see also Table~\ref{tab:MCC_models} below).
Only a purely toroidal initial magnetic field would not evolve in
time, while any magnetic field configuration of initial models labeled
A1 would still stay approximately stationary, since these models
rotate almost rigidly, and thus the initial magnetic field cannot
wind up itself strongly.


\subsubsection{Presupernova models from stellar evolution calculations}

As initial models for the simulations where we use a microphysical EOS
and deleptonization, we employ the solar-metallicity $ 20 \, M_\odot $
(zero-age main sequence) model of \citet{woosley_02_a} (labeled as
model s20 in Table~\ref{tab:CC_models}). On this spherically symmetric
model, which is initially not in equilibrium as it has a non-zero radial
velocity profile, we impose the rotation law~(\ref{rotation_law}),
using the same rotation nomenclature as for the previously described
polytropes in equilibrium.

In addition, we perform calculations with the ``rotating''
presupernova model E20A of \citet{heger00_a}, which we map onto our
computational grids under the assumption of constant rotation on
cylindrical shells of constant $ \varpi $.


\section{Treatment of matter during the evolution}
\label{eos}

In this work we improve upon preceding relativistic stellar core
collapse simulations by using an advanced description of microphysics
as presented in \citet{ott06a} and \citet{dimmelmeier_07_a}. For
comparison to previous results, we also perform simulations with the
simplified hybrid EOS \citep{janka_93_a}. In the following, we
describe both approaches for the treatment of matter.


\subsection{Hybrid EOS}
\label{hybrid_eos}

For calculations employing polytropes in rotational equilibrium as
initial models, we utilize the hybrid EOS. Here the pressure consists
of a polytropic part, $ P_\mathrm{p} = K \rho^\gamma $, with
$ K = 4.897 \times 10^{14} $ (in cgs units), plus a thermal part,
$ P_\mathrm{th} = \rho \epsilon_\mathrm{th} (\gamma_\mathrm{th} - 1) $,
where $ \epsilon_\mathrm{th} = \epsilon - \epsilon_\mathrm{p} $
and where we set $ \gamma_\mathrm{th} = 1.5 $. The thermal
contribution is chosen to take into account the rise of thermal energy
due to shock heating. As $ \rho $ reaches nuclear density at
$ \rho_\mathrm{nuc} = 2.0 \times 10^{14} \mathrm{\ g\ cm}^{-3} $,
$ \gamma $ is raised to $ \gamma_2 = 2.5 $ and $ K $ adjusted
accordingly to maintain monotonicity of $ P $ and $ \epsilon $. Due to
this stiffening of the EOS the core undergoes a so-called
pressure-supported bounce. More details of the hybrid EOS can be found
e.g.\ in \citet{dimmelmeier_02_a}.


\subsection{Microphysical EOS, deleptonization scheme, and neutrino
  pressure}

In our more realistic calculations, for which the models s20 and E20A
from stellar evolution are taken as initial models, we employ the
tabulated non-zero temperature nuclear EOS by \citet{shen_98_a}
in the variant of \citet{marek_05_a} which includes baryonic,
electronic, and photonic pressure components. This gives the fluid
pressure $ P $ (and additional thermodynamic quantities) as a function
of $ \rho $, the temperature $ T $, and the electron fraction $ Y_e $.
Since the code operates with the specific internal energy $ \epsilon $
instead of the temperature $ T $, we determine the corresponding value
for $ T $ iteratively with a Newton--Raphson scheme.

To determine the evolution of $ Y_e $, the state vector, flux vector,
and source vector for the conservation
equations~(\ref{eq:hydro_conservation_equation}), as given in
Eqs.~(\ref{eq:state_vector}--\ref{eq:source_vector}) have to be
augmented by the components
\begin{equation}
  D Y_e,
  \qquad
  D Y_e \hat{v}^i,
  \qquad
  S_{Y_e},
\end{equation}
respectively. The source term $ S_{Y_e} $ is a consequence of the
electron captures during collapse, which reduces $ Y_e $. This
deleptonization also effectively decreases the size of the
homologously collapsing inner core, and has thus a direct influence on
the collapse dynamics and the gravitational wave signal. Hence, it is
essential to include (at least an approximate scheme for)
deleptonization during collapse.

Since multi-dimensional radiation hydrodynamics calculations in
general relativity are not yet computationally feasible, in the
simulations using the microphysical EOS we make use of a a recently
proposed scheme \citep{liebendoerfer_05_a} where deleptonization is
parametrized based on data from detailed spherically symmetric
calculations with Boltzmann neutrino transport. As in
\citet{dimmelmeier_07_a} we take the latest available electron capture
rates \citep{langanke_00_a}, which result in lower values for $ Y_e $
in the inner core at bounce compared to recent results \citep{ott06a}
where standard capture rates were used \citep{rampp_00_a}. Following
\citet{liebendoerfer_05_a}, deleptonization is stopped at core bounce
(i.e.\ as soon as the specific entropy $ s $ per baryon exceeds
$ 3 k_\mathrm{B} $). After core bounce $ Y_e $ is only passively
advected, neglecting any further deleptonization in the nascent PNS.

Neutrino pressure is included only in the regime which is optically
thick to neutrinos, which we define for $ \rho $ being above the
trapping density
$ \rho_\mathrm{t} = 2 \times 10^{12} \mathrm{\ g\ cm}^{-3} $.
Following \citet{liebendoerfer_05_a}, here we approximate the
contribution of the neutrino pressure $ P_\nu $ as an ideal Fermi gas
and include the radiation stress via additional source terms in the
momentum and energy equations for the fluid.


\section{Outline of the numerical approach}
\label{numerics}

The GRMHD numerical code we use in our simulations is
based on the purely hydrodynamic code described in
\citet{dimmelmeier_02_a, dimmelmeier_02_b}, and on its extension
discussed in \citet{cerda05}. It has been described in detail in a
previous paper \citep{nfnr}, which allows us to provide here only
succint information. The code performs the coupled time evolution of
the equations governing the dynamics of the spacetime, the fluid, and
the magnetic field in general relativity. The equations are
implemented in the code using spherical polar coordinates
$ \{ t, r, \theta, \varphi \} $, assuming axisymmetry with respect to
the rotation axis and equatorial plane symmetry at
$ \theta = \pi / 2 $.


\subsection{The hydrodynamics solver}
\label{sec:hydro_solver}

For the evolution of the matter fields we utilize a high-resolution
shock-capturing (HRSC) scheme, which numerically integrates the subset
of equations in system~(\ref{eq:hydro_conservation_equation}) that
corresponds to the purely hydrodynamic variables ($ D $, $ S_i $,
$ \tau $). HRSC methods ensure the numerical conservation of
physically  conserved quantities and a correct treatment of
discontinuities such as shocks \citep[see e.g.][for a review and
references therein]{font_03_a}. We have implemented in the code
various cell-reconstruction procedures, either second-order or
third-order accurate in space, namely minmod, MC, and PHM
\citep[see][for definitions]{toro99}. The time update of the state
vector $ \mb{U} $ is done using the method of lines in combination
with a second-order accurate Runge--Kutta scheme. The numerical fluxes
at the cell interfaces are obtained  using  either the HLL
single-state solver of \citet{harten83} or the symmetric scheme of
\citet{kt00} (KT hereafter). Both solvers yield results with an
accuracy comparable to complete Riemann solvers (with the full
characteristic information), as shown in simulations involving purely
hydrodynamic special relativistic flows \citep{lucas04} and general
relativistic flows in dynamical spacetimes
\citep{shibata_font05}. Tests of both solvers in GRMHD have been
reported recently by \citet{anton06}.


\subsection{Evolution of the magnetic field}
\label{b_field_evolution}

The evolution of the magnetic field needs to be performed in a way
that is different from the rest of the conservation equations, since
the physical meaning of the corresponding conservation equation is
different. Although the induction equation can be written in a flux
conservative way, a supplementary condition for the magnetic field has
to be given (the divergence constraint), which has to be fulfilled at
each time iteration. The physical meaning of these two equations is
the conservation of the magnetic flux in a close volume, in our case
each numerical cell. Therefore, an appropriate numerical scheme has to
be used which takes full profit of such a conservation law. Among the
numerical schemes that satisfy this property \citep[see][for a
review]{toth00}, the constrained transport (CT) scheme \citep{evans88}
has proved to be adequate to perform accurate simulations of
magnetized flows. Our particular implementation of this scheme
\citep[see][for details]{nfnr} has been adapted to the spherical polar
coordinates used in the code. The discretized evolution equations for
the poloidal components of the magnetic field read
\begin{eqnarray}
  \partial_t B^{*\,r} _{i+\half\, j} & = &
  \frac{[\sin \theta \, E^*_\varphi]_{i+\half \, j+\half} -
  [\sin \theta \, E^*_\varphi]_{i+\half \, j-\half}}
  {r_{i+\half \, j} \, \Delta (\cos \theta)_j},
  \label{eq:ct_r}
  \\
  \partial_t B^{*\,\theta}_{i \, j+\half} & = &
  2 \, \frac{[r \, E^*_\varphi]_{i+\half \, j+\half} -
  [r \, E^*_{\varphi}]_{i-\half \, j+\half}}{\Delta r^2_i},
  \label{eq:ct_theta}
\end{eqnarray}
where (in vectorial form) $ \mb{E}^* = \mb{v}^* \times \mb{B}^* $, and
where cell centers are located at $ (i\,j) $, radial interfaces at
$ (i+\half\,j) $, angular interfaces at $ (i\,j+\half) $, and cell
corners (cell edges along the $ \varphi $-direction) at
$ (i+\half\,j+\half) $. We note that the evolution equation for the
toroidal magnetic field is analog to that used for the hydrodynamics,
since in axisymmetry this component does not play any role in the CT
scheme. The previous expressions are used in the numerical code to
update the magnetic field. The only remaining aspect is to give an
explicit expression for the value of $ E^*_i $. A practical way to
calculate $ E^*_\varphi $ from the numerical fluxes in the adjacent
interfaces \citep{balsara99} is
\begin{eqnarray}
  E^*_{\varphi \, i+\half \, j+\half} & = &
  - \frac{1}{4} \left[ (F^r)^\theta_{i \, j+\half} +
  (F^r)^\theta_{i+1 \, j+\half} \right.
  \nonumber
  \\ & & \left. \qquad
  - (F^\theta)^r_{i+\half \, j} -
  (F^\theta)^r_{i+\half \, j+1} \right],
\end{eqnarray}
where the fluxes~(\ref{eq:flux_vector}) are obtained in the usual way
by solving the Riemann problem at the interfaces. The combination of
the CT scheme and this way of computing the electric field is called
the flux-CT scheme. It is used in all numerical simulations reported
in this paper. Finally, the time discretization of
Eqs.~(\ref{eq:ct_r}) and~(\ref{eq:ct_theta}) is performed in the same way
as for the fluid evolution equations.


\subsection{The metric solver}

The CFC metric equations are five nonlinear elliptic coupled
Poisson-like equations which can be written in compact form as
$ \hat{\Delta} \mb{u} (\mb{x}) = \mb{f} (\mb{x}; \mb{u} (\mb{x}))$,
where $ \mb{u} = u^k = (\phi, \alpha \phi, \beta^j) $, and
$ \mb{f} = f^k $ is the vector of the respective sources. These five
scalar equations for each component of $ \mb{u} $ couple to each other
via the source terms that in general depend on the various components
of $ \mb{u} $. We use a fix-point iteration scheme in combination with
a linear Poisson solver to solve these equations. Further details on
this type of metric solver can be found in \citet{cerda05} and
\citet{dimmelmeier_02_a}.


\subsection{Setup of the numerical grid}

\begin{table*}[t!]
  \centering
  \caption{Hydrodynamical and magnetic field properties of all models
    computed in this work. From left to right the columns report the
    name of the model, the stationarity properties of the initial
    magnetic field, the maximum rest-mass density
    $ \rho_\mathrm{max} $, the maximum poloidal field
    $ |B_\mathrm{polo}|_\mathrm{max} $, the maximum toroidal field
    $ |B_\varphi|_\mathrm{ max} $, the rotational energy parameter
    $ \beta_\mathrm{rot} $, the magnetic energy parameter
    $ \beta_\mathrm{mag} $, its poloidal contribution
    $ \beta_\mathrm{polo} $, and the central angular velocity
    $ \Omega_\mathrm{c} $. The time scale $ \tau_\Omega $ of the
    growth of the magnetic field by the $ \Omega $-dynamo and the
    saturation time $ t_\mathrm{sat} $ for this process are also
    shown. }
  \begin{tabular}{l|c|c@{~~}c@{~~}ccccc|rr}
    \hline \hline
    Model
    & stationary
    & $ \rho_\mathrm{max} $
    & $ |B_\mathrm{polo}|_\mathrm{max} $
    & $ |B_{\varphi}|_\mathrm{ max} $
    & $ \beta_\mathrm{rot} $
    & $ \beta_\mathrm{mag} $
    & $ \beta_\mathrm{polo} $
    & $ \Omega_\mathrm{c} $
    & \multicolumn{1}{c}{$ \tau_\Omega $}
    & \multicolumn{1}{c}{$ t_\mathrm{sat} $}
    \\
    &
    & $ \displaystyle \left[ 10^{14} \frac{\mathrm{g}}{\mathrm{\ cm}^3} \right] $
    & [$ 10^{10} \mathrm{\ G} $]
    & [$ 10^{10} \mathrm{\ G} $]
    & [$ 10^{-2} $]
    & [$ 10^{-8} $]
    & [$ 10^{-8} $]
    & [$ \mathrm{ms}^{-1} $]
    & \multicolumn{1}{c}{[s]}
    & \multicolumn{1}{c}{[s]}
    \\ [0.5 em]
    \hline
    A2B4G1-D3M0  & no      & 0.47 &  \z400 &  1467 &  15.6 & \z8.3 & 0.9 & 0.36 &  85 & 10.3 \\
    A1B3G3-D3M0  & approx. & 4.22 &   1719 &  2522 & \z2.3 & \z1.2 & 0.8 & 3.96 &   7 &  0.3 \\
    A1B3G3-T3M0  & yes     & 4.22 & \zz\z0 &  1714 & \z2.3 & \z0.2 & 0.0 & 3.96 & --- &  --- \\
    A1B3G5-D3M0  & approx. & 4.57 &   1146 &  1275 & \z0.9 & \z0.5 & 0.4 & 3.91 &  21 &  0.7 \\
    A1B3G5-T3M0  & yes     & 4.57 & \zz\z0 &  1542 & \z0.9 & \z0.2 & 0.0 & 3.91 & --- &  --- \\
    A1B3G5-DT3M0 & approx. & 4.57 &   1146 &  1537 & \z0.9 & \z0.6 & 0.4 & 3.91 &  21 &  0.6 \\
    A3B3G5-D3M0  & no      & 3.73 &  \z984 &  1672 & \z2.3 & \z0.6 & 0.4 & 3.75 &  24 &  1.1 \\
    A4B5G5-D3M0  & no      & 1.74 &   1094 &  2716 & \z8.5 & \z4.4 & 1.2 & 1.18 &  24 &  2.1 \\
    A4B5G5-T3M0  & yes     & 1.74 & \zz\z0 &  1626 & \z8.5 & \z0.4 & 0.0 & 1.18 & --- &  --- \\
    s20A1B1-D3M0 & approx. & 2.69 &   1221 & \z162 & \z0.6 & \z2.9 & 2.9 & 1.34 &  22 &  0.5 \\
    s20A2B2-D3M0 & no      & 2.75 &   1849 &  3574 & \z5.8 & \z7.6 & 3.2 & 3.55 &   7 &  0.5 \\
    s20A2B2-T3M0 & yes     & 2.75 & \zz\z0 &  1365 & \z5.8 & \z0.9 & 0.0 & 3.55 & --- &  --- \\
    s20A1B5-D3M0 & approx. & 2.69 &   1100 &  1011 & \z7.8 & \z3.2 & 2.4 & 3.89 &  10 &  0.8 \\
    s20A1B5-T3M0 & yes     & 2.69 & \zz\z0 &  1447 & \z7.8 & \z1.2 & 0.0 & 3.89 & --- &  --- \\
    E20A-D3M0    & no      & 2.29 &   2343 &  7503 & \z7.7 &  23.4 & 1.8 & 4.37 &   6 &  0.5 \\
    E20A-T3M0    & yes     & 2.29 & \zz\z0 &  2739 & \z7.7 & \z1.9 & 0.0 & 4.37 & --- &  --- \\
    \hline \hline
  \end{tabular}
  \label{tab:MCC_models}
\end{table*}

We perform all axisymmetric simulations with a resolution
($ n_r \times n_\theta $) of $ 300 \times 30 $ zones, except for
models labeled A4B5G5 in which a resolution of $ 300 \times 40 $ is
used due to the more complex angular structure. In both cases the
radial grid is equally spaced for the first 100 points, with a grid
spacing of $ 100 \mathrm{\ m} $. The remaining radial zones are
logarithmically distributed to cover the outer parts of the star and
the exterior artificial low-density atmosphere. The angular grid is
equally spaced and we assume equatorial symmetry. We have performed
resolution tests and we have found that
such a resolution is adequate for our simulations \citep[see][for 
details]{cerda_phd,nfnr}. As a consequence of our various code tests (see
Appendix~\ref{app:tests}) all results discussed in Sect.~\ref{results}
correspond to simulations performed using PHM reconstruction and the
HLL solver for the hydrodynamics.


\section{Results}
\label{results}

We now present the main results from our numerical simulations of
rotational magnetized core collapse to neutron stars. First, we note
that a quantitative summary of our findings is reported in
Table~\ref{tab:MCC_models}, to which we will refer repeteadly. The
dynamics of the models we have selected is identical to the dynamics
of the unmagnetized ones, since the passive field approximation is
used. Therefore, we will not describe here all the morphological features
of the hydrodynamics of both models with the hybrid EOS (simplified
models hereafter) and models with the microphysical EOS and the
deleptonization scheme (microphysical models hereafter), as they have
been discussed in detail in \citet{dimmelmeier_02_a, dimmelmeier_02_b},
and \citet{ott06b} as well as \citet{dimmelmeier_07_a,
  dimmelmeier_07_b}, respectively. (It is worth to emphasize, however,
the excellent agreement found in the hydrodynamical simulations
performed with three independent numerical codes.) We pay more
attention instead to the magnetic field evolution. In all our
simulations an initial magnetic field strength of
$ B^*_0 = 10^{10} \mathrm{\ G} $ is considered. This value is an upper
limit for the T3M0 models, since the expected initial toroidal
magnetic field is of this order \citep{heger05}. However, for the D3M0
models, this field strength is already at (or above) the upper end of
the astrophysically expected values.

For all models we first present results for identical values of
$ B^*_0 $, in a way that we can study the different effects and
compare them properly. Afterwards we present the results scaled to
lower, astrophysically expected values. We anticipate that our results
can change if some of the several assumptions made in our simulations
(axisymmetry and passive field approximation) are relaxed. An estimation
and discussion of these effects can be found in Sect.~\ref{sub:ampli}.


\subsection{Evolution of the magnetic energy parameter}
\label{sec:beta_evol}

\begin{figure}[t!]
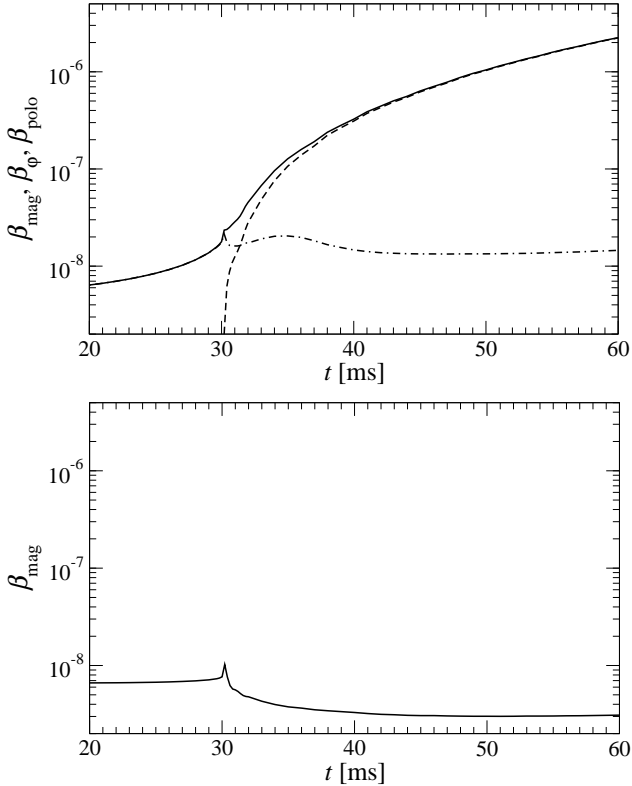

  \centering
  \resizebox{0.47\textwidth}{!}{\includegraphics*{7432f2a.eps}}
  \\ [0.5 em]
  \resizebox{0.47\textwidth}{!}{\includegraphics*{7432f2b.eps}}
  \caption{Time evolution of the magnetic energy parameters
    $ \beta_\mathrm{mag} $ (solid line), $ \beta_\varphi $ (dashed
    line), and $ \beta_\mathrm{polo} $ (dashed-dotted line) for models
    A1B3G5-D3M0 (top panel) and A1B3G5-T3M0 (bottom panel).}
  \label{fig:beta_mcollapse}
\end{figure}

\begin{figure}[t!]
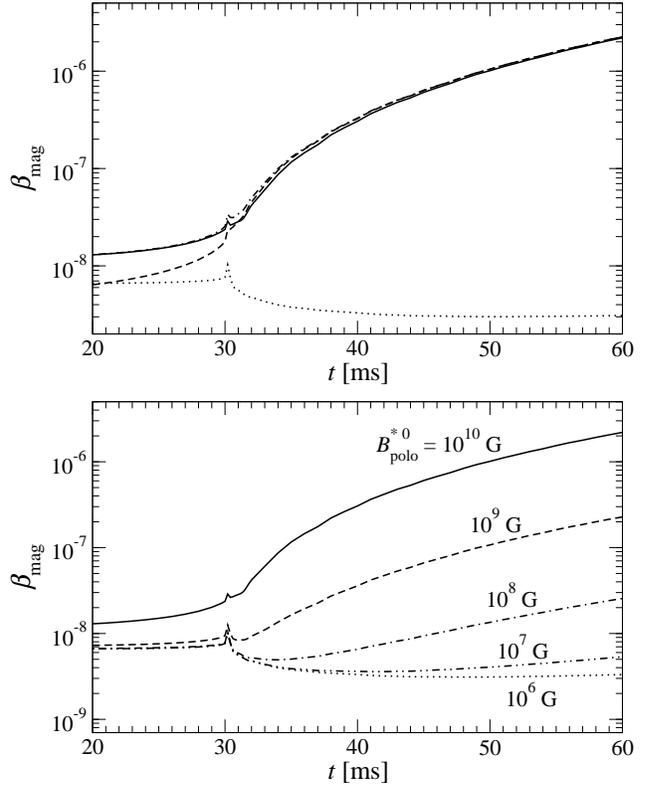

  \centering
  \resizebox{0.47\textwidth}{!}{\includegraphics*{7432f3a.eps}}
  \\ [0.5 em]
  \resizebox{0.47\textwidth}{!}{\includegraphics*{7432f3b.eps}}
  \caption{Time evolution of the magnetic energy parameter
    $ \beta_\mathrm{mag} $. The top panel shows $ \beta_\mathrm{mag} $
    for model A1B3G5-DT3M0 (solid line) and the comparison with the
    composition (dot-dashed line) of models A1B3G5-D3M0 (dashed line)
    and A1B3G5-T3M0 (dotted line). The bottom panel shows the
    evolution of $ \beta_\mathrm{mag} $ for the same hydrodynamic
    model (A1B3G5) with an initial value of the toroidal field of
    $ B^{*\,0}_\varphi = 10^{10} \mathrm{\ G} $, and varying values of
    the initial poloidal field $ B^{*\,0}_\mathrm{polo} $, from
    $ 10^{6} $ to $ 10^{10} \mathrm{\ G} $.}
  \label{fig:beta_mcollapse2}
\end{figure}

The evolution of the energy parameter for the magnetic field
$ \beta_\mathrm{mag} = E_\mathrm{mag} / |E_\mathrm{b}| $ can be seen
in Fig.~\ref{fig:beta_mcollapse} for model A1B3G5 of our sample. In
order to analyze the amplification of the magnetic field, we separate
the effects of the different components of the magnetic field into
$ \beta_\varphi $ for the toroidal component and $ \beta_\mathrm{polo} =
\beta_\mathrm{mag} - \beta_\varphi $ for the poloidal component, which
are also plotted in the figure. As the collapse proceeds the magnetic
field grows by at least two reasons: First, the radial flow compresses
the magnetic field lines, amplifying the existing poloidal and
toroidal magnetic field components. Second, during the collapse of a
rotating star differential rotation is produced and increased, even
for rigidly rotating initial models (see
e.g.\ \citet{dimmelmeier_02_b}). Hence, if a seed poloidal field
exists, the $ \Omega $-dynamo mechanism winds up the poloidal field
lines into a toroidal component. This (linear) amplification process
generates a toroidal magnetic field component, even from purely
poloidal initial configurations. The toroidal component of the
magnetic field is affected by the two effects, while the poloidal
field is only amplified by radial compression of the field lines.
Thus, even if the initial magnetic field configuration is purely
poloidal, the toroidal component dominates after some dynamical time.
To study the differences in the evolution of the magnetic field
depending on the initial magnetic field we now describe in detail the
features of model A1B3G5 with different initial magnetic field
configurations.

In model A1B3G5-D3M0 the initial magnetic field is entirely poloidal.
The top panel of Fig.~\ref{fig:beta_mcollapse} shows that
$ \beta_\varphi $ (dashed line) grows much faster than
$ \beta_\mathrm{polo} $, particularly after bounce
($ t \sim 30 \mathrm{\ ms} $) when the radial compression mechanism
stops. We note that the magnetic field considered is weak enough
not to affect the dynamics, with the final $ \beta_\mathrm{mag} \ll 1 $.

If we consider a purely toroidal magnetic field initially, as model
A1B3G5-T3M0, the only amplification mechanism present in our
simulations is the radial compression, since no poloidal field can be
wound up. The bottom panel of Fig.~\ref{fig:beta_mcollapse} shows the
behaviour of $ \beta_\mathrm{mag} $ for model A1B3G3-T3M0. It is
important to notice that during the collapse $ \beta_\mathrm{mag} $
hardly grows (for other models of the T3M0 series it even decreases)
since the radial compression is a very inefficient mechanism to
amplify the magnetic field. As a result, for some models the final PNS
is ``less magnetized'' than the progenitor core in the sense that
$ \beta_\mathrm{mag} $ at bounce is smaller than it is before the
collapse. We note that the evolution of this kind of purely toroidal
models could change completely if the axisymmetry condition were
removed, since in three dimensions there are mechanisms that can
transform a toroidal magnetic field into a poloidal one. Some of these
mechanisms are discussed in Sect.~\ref{sub:ampli} below.

To check whether the ``composition rule'' (see
Sect.~\ref{sec:passive}) is valid we consider next a mixed
configuration of poloidal and toroidal magnetic fields at the beginning
of the simulation (model A1B3G5-DT3M0). The top panel of
Fig.~\ref{fig:beta_mcollapse2} shows with a solid line the time
evolution of $ \beta_\mathrm{mag} $ for model A1B3G5-DT3M0 and with a
dot-dashed line the composition of the individual values for
$ \beta_\mathrm{mag} $ in models A1B3G5-D3M0 and A1B3G5-T3M0
with identical initial field strengths. (The separate evolutions for
the latter are also included in the plot as dashed and dotted lines,
respectively.) The agreement of the two evolution paths is remarkable,
which shows that the ``composition rule'' works properly for our
simulations. Therefore, we can use it to obtain any desirable
composition of magnetic fields with a single hydrodynamic evolution of
the two models D3M0 and T3M0. For the particular composition showed in
this model, the final value of $\beta_\mathrm{mag}$ depends very
weakly on the initial toroidal magnetic field component. In other
words, the structure of the magnetic field of the PNS will depend
almost exclusively on the radial compression of the initial poloidal
component of the magnetic field.

Next, we consider a ``composition'' of these models with different
initial magnetic field strength. We keep the initial toroidal
component fixed at a realistic value,
$ B^{*\,0}_\varphi = 10^{10} \mathrm{\ G} $, and change the initial
poloidal component in a range that spans from
$ B^{*\,0}_\mathrm{polo} = 10^{10} \mathrm{\ G} $ down to the
astrophysically more realistic value of $ 10^6 \mathrm{\ G} $. The
bottom panel of Fig.~\ref{fig:beta_mcollapse2} shows the time
evolution of $ \beta_\mathrm{mag} $ for these different
configurations. For lower values of $ B^{*\,0}_\mathrm{polo} $, the
$ \Omega $-dynamo mechanism becomes increasingly slower and the
initial toroidal component becomes important for the magnetic field
configuration of the PNS. In the lowest initial poloidal field case
analyzed, the magnetic field of the PNS is completely toroidal and
depends exclusively on the initial magnetic field configuration.

\begin{figure}[t!]
  \centering
  \resizebox{0.475\textwidth}{!}{\includegraphics*{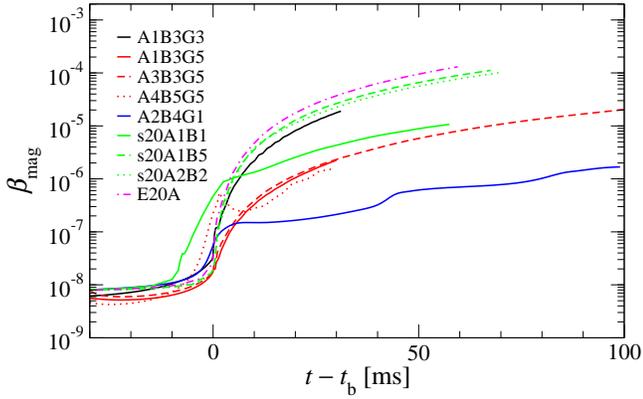}}
  \caption{Time evolution of the magnetic energy parameter
    $\beta_\mathrm{mag}$ for all simulated models  with initial purely
    poloidal magnetic field (label D3M0).}
  \label{fig:beta_mcollapse_comp}
\end{figure}

\begin{figure}[t!]
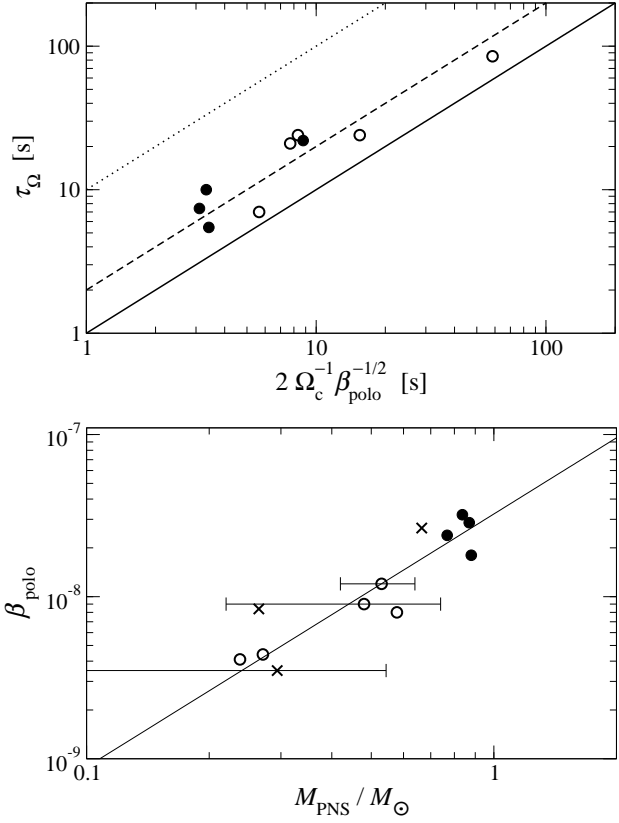

  \centering
  \resizebox{0.47\textwidth}{!}{\includegraphics*{7432f5a.eps}} \\
  [0.5 em]
  \resizebox{0.47\textwidth}{!}{\includegraphics*{7432f5b.eps}}
  \caption{Top panel: relation between the fitted values for the
    growth time $ \tau_\Omega $ of the $ \Omega $-dynamo and the upper
    limit given by
    $ 2 \, \Omega_\mathrm{c}^{-1} \beta_\mathrm{polo}^{-1/2} $. The
    solid line represents the upper limit, the dashed line corresponds
    to 50\% of that limit, and the dotted line to 10\%. Bottom panel:
    magnetic energy parameter $ \beta_\mathrm{polo} $ of the poloidal
    component of the magnetic field at bounce versus the average mass
    $ M_\mathrm{PNS} $ of the PNS after bounce, and its fit to a power
    law. Error bars are shown for models with mass variations larger
    than 5\% after bounce. Both microphysical models (filled circles)
    and simplified models (open circles) are presented. Newtonian models
    (crosses) are also shown, but are not used in the fit.}
  \label{fig:beta_mcollapse_comp2}
\end{figure}

The remaining computed models of our sample, including those with
microphysics, behave qualitatively in a very similar manner, although
quantitative differences can be found in the amplification of the
magnetic field during the collapse, and the amplification rates after
bounce due to the $ \Omega $-dynamo.
Fig.~\ref{fig:beta_mcollapse_comp} shows the evolution of the magnetic
energy parameter $ \beta_\mathrm{mag} $ for all the simulated models
with initial purely poloidal magnetic field (label D3M0). For all
models we find the following relation between the collapse time and
the amplification rate of the magnetic field after bounce
\citep[which, however, does not hold for model A2B4G1-D3M0, as this is
the only model of our sample for which the collapse is halted not by
the stiffening of the EOS, but rather by centrifugal forces at
subnuclear densities; cf.][]{dimmelmeier_02_b}: Models with large
collapse times, such as all microphysical models as well as the
simplified model A1B3G3-D3M0, exhibit a more efficient amplification
of the magnetic field as compared to the rapid collapse models (G5
series). To quantify the differences between the models we estimate
the time scale $ \tau_\Omega $ for the amplification of the magnetic
field by fitting the post-bounce evolution of $ \beta_\mathrm{mag} $
to
\begin{equation}
  \beta_\mathrm{mag} = \left( \frac{t}{\tau_\Omega} \right)^2.
\end{equation}
The resulting values can be found in Table~\ref{tab:MCC_models}. The
time scale should depend on the central angular velocity
$ \Omega_\mathrm{c} $ of the PNS, and on the strength of the poloidal
magnetic field that can be wound up, which can be estimated from
$ \beta_\mathrm{polo} $. Hence, the following expression should be
valid in the most efficient scenario (see Appendix~\ref{app:odynamo}
for details):
\begin{equation}
  \tau_\Omega = \frac{2}{\Omega_\mathrm{c} \sqrt{\beta_\mathrm{polo}}}.
  \label{upper_limit}
\end{equation}
To check this relation we plot in the top panel of
Fig.~\ref{fig:beta_mcollapse_comp2} the value of the fit for
$ \tau_\Omega $ versus the value from the previous analytic
expression. Apparently for all models the growth time of the
$ \Omega $-dynamo is always larger than that of the most efficient
situation (solid line in the figure), and corresponds to a fraction
(30\%\,--\,90\%) of the upper limit~(\ref{upper_limit}). This relation
shows that in order to obtain higher amplification rates of the
magnetic field not only strong rotation is needed, but also a
sufficient compression of the poloidal magnetic field during the
collapse.

Furthermore, we also find a relation between the value of $
\beta_\mathrm{polo} $ and the mass enclosed in the neutrino
sphere\footnote{In all models, we define the neutrino sphere as the
  surface inside the core where the density equals the trapping density
  $ \rho_\mathrm{trap} = 2 \times 10^{12} \mathrm{\ g\ cm}^{-3} $. In
  the microphysical models, above this density neutrinos are assumed to
  be trapped in the medium \citep{liebendoerfer_05_a}.},
$ M_\mathrm{PNS} $ hereafter (see bottom panel of
Fig.~\ref{fig:beta_mcollapse_comp2}). Since most of the magnetic field
lines compressed by the collapse are located inside the neutrino
sphere, it is easy to understand that more massive PNS have higher
magnetic energies. The fit to a power law of the data shown in
Fig.~\ref{fig:beta_mcollapse_comp2} yields 
\begin{eqnarray}
  \beta_\mathrm{polo} & = & (3.2 \pm 0.5) \times
  \nonumber
  \\
  & & 10^{-8} 
  \left( \frac{M_\mathrm{PNS}}{M_\odot} \right)^{(1.6 \pm 0.2)}
  \left( \frac{B^*_0}{10^{10} \mathrm{\ G}} \right)^2.
  \label{eq:beta_mcollapse_mass}
\end{eqnarray}
As discussed in detail by \citet{dimmelmeier_07_a, dimmelmeier_07_b},
in the microphysical models the mass of the homologously collapsing
inner core at bounce has a value of $ \sim 0.5 \, M_\odot $ (for the
rotation rates considered here). This is also consistent with the high
mass $ M_\mathrm{PNS} \sim 0.8 \, M_\odot $ of the PNS in these
models, as shown in Fig.~\ref{fig:beta_mcollapse_comp2}. To obtain
masses in this range in models with a simple matter treatment, the
adiabatic index would require a value $ \gamma_1 \gtrsim 1.32 $,
which is close to $ 4 / 3 $. Already for moderate rotation, this choice
would cause the core to undergo multiple centrifugal bounces at
densities lower than nuclear density (as exemplified here in model
A2B4G1), which is a dynamical behavior that does not occur at all in
microphysical models \citep[][see also the related discussion in
Sect.~\ref{gravitational_waves}]{dimmelmeier_07_a, dimmelmeier_07_b}.
Therefore, only the microphysical models feature a collapse to a PNS
that has both high densities and is in addition comparably heavy. This
combination, which cannot be realized with the simplified models,
explains the higher growth rates of the magnetic field due to the
$ \Omega $-dynamo observed if improved microphysics is taken into
account.

Combining Eqs.~(\ref{upper_limit}) and~(\ref{eq:beta_mcollapse_mass})
we can establish an upper limit to the growth rate of the magnetic
field due to the $ \Omega $-dynamo using only hydrodynamic quantities,
namely $ \Omega_\mathrm{c} $ and $ M_\mathrm{PNS} $, and the strength
of the magnetic field in the progenitor, $ B^*_0 $. This limit is
given by 
\begin{eqnarray}
  \tau_{\Omega} & = & (11.18 \pm 0.9) \times
  \nonumber
  \\
  & & \left( \frac{1 \mathrm{\ ms}^{-1}}{\Omega_\mathrm{c}} \right)
  \left( \frac{M_\odot}{M_\mathrm{PNS}} \right)^{(0.8 \pm 0.1)}
  \left( \frac{10^{10} \mathrm{\ G}}{B^*_0} \right) \mathrm{\ s}.
\end{eqnarray}
This relation can be very useful to estimate how fast the magnetic
field grows in a collapsed star, under the assumption of a weak
magnetic field and with a similar poloidal configuration in the
progenitor, using data from purely hydrodynamical simulations (with no
magnetic fields). As a proof of consistency and in order to assess the
quality of this estimate we have computed $ \tau_\Omega $ with this
method. We find that in all cases the estimate is a lower limit for
the actual value of $ \tau_\Omega $ obtained from the numerical
simulations and deviates by at most 30\%.


\subsection{Convection}
\label{subsec:convection}

One of the most important features that can affect the evolution of
the magnetic field in stellar core collapse to a PNS is the presence
of convection. We present here a detailed analysis of this effect in our
simulations. Since in all of our models the magnetic field is weak,
the discussion can be performed without considering its influence. We
also note that due to the approximations made in our simulations,
specifically the lack of a consistent neutrino transport scheme, our
findings regarding convection should not be considered as definite.

The stability conditions for a rotating star are given by the so-called
Solberg--H{\o}iland criteria \citep{tassoul78},
\begin{equation}
  \begin{array}{rcl}
    \mathcal{C}_\mathrm{SH1} & = & \mb{g} \cdot {\bf \mathcal{B}} +
    \mathcal{J} \cdot \nabla\varpi > 0,
    \\ [0.2 em]
    \mathcal{C}_\mathrm{SH2} & = & (\mb{g} \times \nabla \varpi) 
    (\mathcal{B} \times \mathcal{J}) > 0,
  \end{array}
  \label{eq:shcriteria}
\end{equation}
where $ \mb{g} $ is the gravitational acceleration, and the buoyancy
and rotational terms are respectively given by
\begin{equation}
  \mathcal{B} = \frac{\nabla \rho}{\rho} - \frac{\nabla P}{P \Gamma_1},
  \qquad
  \mathcal{J} = \frac{1}{\varpi^3} \nabla (\Omega^2 \varpi^4),
\end{equation}
with $ \Gamma_1 = (\partial \ln P / \partial \ln \rho)_{s, Y_e} $.
Note that in the first condition of Eq.~(\ref{eq:shcriteria}),
$ N^2 = \mb{g} \cdot \mathcal{B} $ is the Brunt--V\"ais\"al\"a
frequency and $ \kappa^2 = \mathcal{J} \cdot \nabla \varpi $ is the
epicyclic frequency. In the case of either no rotation or uniform rotation
the Solberg--H{\o}iland criteria reduce to the well known
Schwarzschild criterion, $ N^2 > 0 $. If one of the two conditions is
not satisfied, convective instability  develops. Following
\cite{miralles04}, the time scale of the fastest growing mode can be
computed as
\begin{equation}
  \tau_\mathrm{\,SH} = \imag \left[ \left(
  \frac{\mathcal{C}_\mathrm{SH1}}{2} -
  \frac{1}{2} \sqrt{\mathcal{C}^2_\mathrm{SH1} -
  4 \, \mathcal{C}_\mathrm{SH2}} \right)^{-1/2} \right].
  \label{eq:shtimescale}
\end{equation}
It is very useful to express the buoyancy terms in the
conditions~(\ref{eq:shcriteria}) in terms of the contributions of the
entropy and electron fraction gradients,
\begin{equation}
  \mathcal{B} = \xi \, \nabla s + \delta \, \nabla Y_\mathrm{e},
  \label{eq:buoyancy_reformulation}
\end{equation}
with $ \xi = - \partial \ln P / \partial s|_{\rho, Y_e} / \Gamma_1 $
and $ \delta = - \partial \ln P / \partial Y_e|_{\rho, s} / \Gamma_1 $.
We point out that the Solberg--H{\o}iland criteria are valid exactly
only in Newtonian gravity, and thus we use them here only as 
estimates. In order to assess the influence of general relativistic
corrections, we also evaluate Eq.~(\ref{eq:shcriteria}) using
covariant derivatives with respect to the CFC metric, which yields
very similar results. Note also that the Solberg--H{\o}iland criteria
are based on a local instability analysis, while the convection
observed in our simulations covers extended regions.

\begin{figure}[t!]
  \centering
  \resizebox{0.5\textwidth}{!}{\includegraphics*{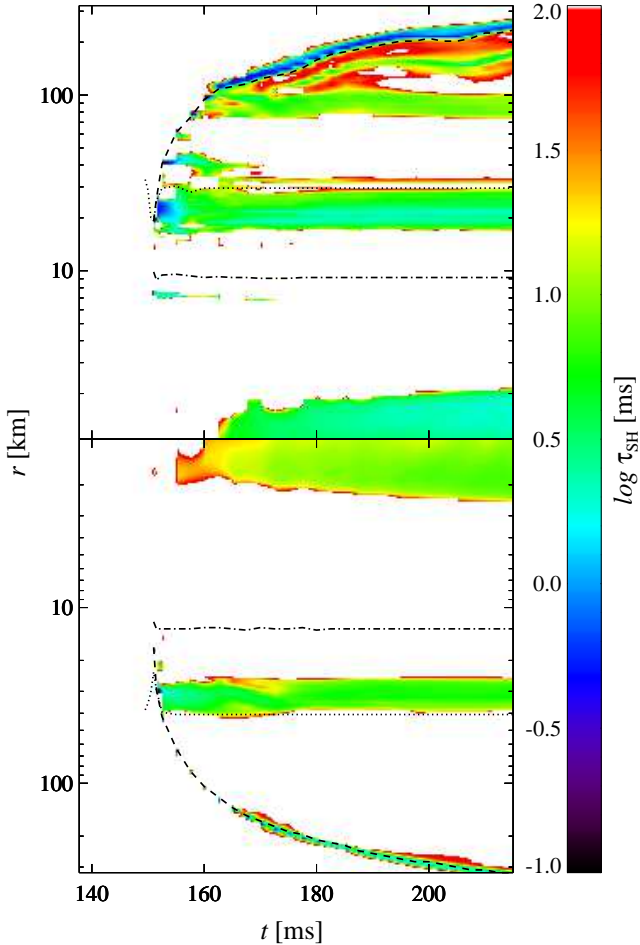}}
  \caption{Time evolution of the angular averaged value of the growth
    time scale $ \tau_\mathrm{\,SH} $, which indicates convectively
    unstable regions according to the Solberg--H{\o}iland criteria.
    White regions represent convectively stable regions. The top half of the
    figure   shows angular averages near the pole ($ 0 < \theta < \pi / 6 $),
    while the bottom half shows the corresponding averages near the 
    equator ($ \pi / 3 < \theta < \pi / 2 $). The shock radius (dashed lines),
    neutrino sphere radius (dotted line), and radius of shock
    formation (dash-dotted line) are also displayed.}
  \label{fig:convection}
\end{figure}

In Fig.~\ref{fig:convection} we show the extent of the convectively
unstable regions according to the Solberg-H{\o}iland
criteria~(\ref{eq:shcriteria}) after core bounce for models of the
series s20A1B5, by plotting the time evolution of angle-averaged
values for the convective growth time scale $ \tau_\mathrm{\,SH} $.
From this figure it becomes apparent that two regions are susceptible
to developing instabilities: the region just below the neutrino sphere
(between about $ 20 \mathrm{\ km} $ and $ 40 \mathrm{\ km} $) and
extended regions behind the shock. The innermost $ 2 \mathrm{\ km} $
of the star are also convectively unstable, but we suspect that the
small negative entropy gradient responsible of this unstable region is
a numerical artifact of the inner boundary, related to the so-called
wall heating effect commonly appearing in shock reflection
experiments~\citep{donat_96_a}. In our simulations of models of the
series s20A1B5, convective motions indeed occur in those unstable
regions as predicted by the instability criteria, as well as in the
surrounding regions due to overshooting. We also find that the time
scale of the onset of the observed instability is correctly estimated
by Eq.~(\ref{eq:shtimescale}).

Below the neutrino sphere ($ 20\mbox{\--\,}40 \mathrm{\ km} $),
convection sets in inmediately after bounce, with typical maximum
velocities of about $ 2 \times 10^4 \mathrm{\ km\ s}^{-1} $. The
velocities progressively decrease until the end of the simulation (at
about $ 65 \mathrm{\ ms} $ after bounce) with average values around
$ 100 \mathrm{\ km\ s}^{-1} $, although convection does not disappear
completely. Behind the shock ($ 100\mbox{\,--\,}200 \mathrm{\ km} $),
the typical convective velocities are of the order of
$ 1000 \mathrm{\ km\ s}^{-1} $, with maximum values in some regions of
$ 10^4 \mathrm{\ km\ s}^{-1} $. This magnitude remains until the end
of the simulation.

For a more detailed analysis we separately evaluate the different
contributions in the Solberg--H{\o}iland
criteria~(\ref{eq:shcriteria}) with $ \mathcal{B} $ in the form of
Eq.~(\ref{eq:buoyancy_reformulation}). Since the radial gradient of
$ Y_e $ is positive (as deleptonization is stronger towards the center
during the collapse), this has an stabilizing effect against
convection. Similarly, rotation also suppresses convection, since the
epicyclic frequency $ \kappa^2 $ is positive everywhere. Convective
instability can thus only appear in regions with a sufficiently large
negative radial entropy gradient. Such a gradient occurs in the region
already swept by the shock front. Shock heating creates entropy most
strongly close to the neutrino sphere at a radius of about
$ 30 \mathrm{\ km} $ (see Fig.~\ref{fig:convection}), producing a
steep gradient there $ 1 \mathrm{\ ms} $ after core bounce. Behind the
shock front, which then propagates to larger radii at lower densities
and decelerates, another region with a negative gradient also
appears. All our microphysical models show very similar qualitative
behavior with some variations due to different angular momentum
distribution and the description of matter.

In models with slower rotation (i.e.\ the s20A1B1 series), strong
convection sets in immediately after the occurence of the negative
entropy gradient close to the neutrino sphere. For models with very
little rotation (which have not been considered in this work), such
convective overturn is strong enough to be clearly visible in the
post-bounce gravitational wave signal \citep{dimmelmeier_07_a,
dimmelmeier_07_b}. Within about $ 20 \mathrm{\ ms} $ after core
bounce, convection has managed to smooth out the entropy gradient
around the neutrino sphere, thus removing the condition for sustained
convection. Accordingly, convection is strongly damped, the vortices
disappear quickly, and the low-frequency contribution to the
gravitational wave signal is no longer visible. This fast convective
transient near the neutrino sphere has been observed in numerical
simulations without any neutrino treatment
\citep[see e.g.][]{Burrows1992,mueller1997}, and also in simulations
using a neutrino diffusion scheme \citep{Swesty2005}, although in the
latter case the time scale for damping of convection is shorter
($ \sim 10 \mathrm{\ ms} $) than in our case. However, in simulations
including state-of-the-art Boltzmann neutrino transport
\citep{mueller_04_a}, a few ms after core bounce no significant
convection remains in this region, and no traces in the gravitational
wave signal can be found. We attribute this disagreement with our
results to the simplified neutrino treatment in our models, which
cannot properly take into account the deleptonization of the PNS after
core bounce. As the deleptonization of the PNS is initially strongest
when the shock travels through the neutrino sphere, we expect the most
significant inaccuracies of our formulation there. We therefore
conclude that the convection over $ \sim 20 \mathrm{\ ms} $, which we
observe in the neutrino sphere region, is an artifact that should
disappear once a more realistic neutrino description is included.

In more rapidly rotating models, the stabilizing effect of rotation in
the Solberg--H{\o}iland criteria prevents the strong transient we
find in the slowly rotating models from developing, and significantly 
weaker convection is present in this region. However, irrespective of
rotation, convection vortices are formed behind the decelerating shock 
front. On post-bounce evolution times of several $ 10 \mathrm{\ ms} $,  
the weak but persistent convection is unable to remove the entropy 
gradient behind the shock, except near the rotation axis, where the specific
angular momentum is smaller, and convection is stronger. 

Rotation also influences the shape of the convective cells. If the
buoyancy terms in the Solberg--H{\o}iland
criteria~(\ref{eq:shcriteria}) are much larger than the rotation
terms, the convective cells show no preferred direction. We observe
this feature particularly in models with slower rotation (the s20A1B1
series), and to a lesser degree also in other convectively unstable
models in the first few ms after bounce. If the buoyancy terms are
comparable in magnitude to the rotation terms, convection develops
preferredly parallel to the rotation axis \citep[see
e.g.][]{miralles04}. This effect is present in our microphysical
models at later phases, as the entropy gradient has already been
partially smoothed out and the buoyancy terms have become smaller.

In contrast to the microphysical models, which show remarkable
convection in the PNS and behind the shock front, models with a
simplified matter treatment exhibit either no convection at all, or
only close to the neutrino sphere (in the case of models of the A1B3G5
series). This is a consequence of using the hybrid EOS in the latter
models, which is unable to properly decelerate the shock after core 
bounce and turn it into an accretion shock. Hence in these models
the entropy gradient is mostly positive behind the shock.


\subsection{Structure of the magnetic field}
\label{subsec:morpho}

\begin{figure*}[t!]
  \centering
  \resizebox{0.98\textwidth}{!}{\includegraphics*{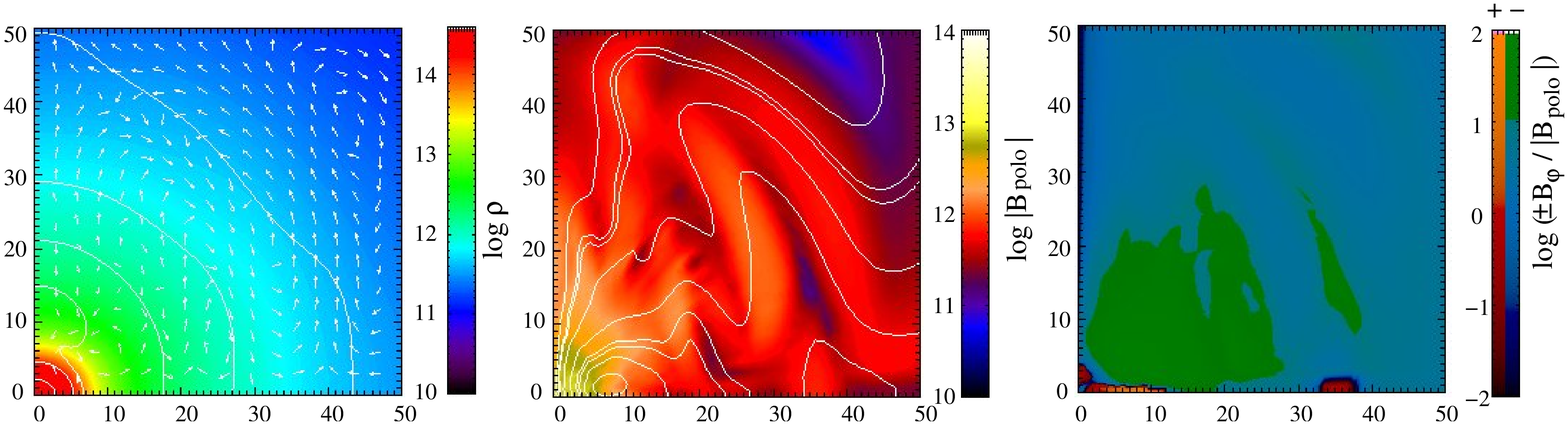}}
  \\ [0.5 em]
  \resizebox{0.98\textwidth}{!}{\includegraphics*{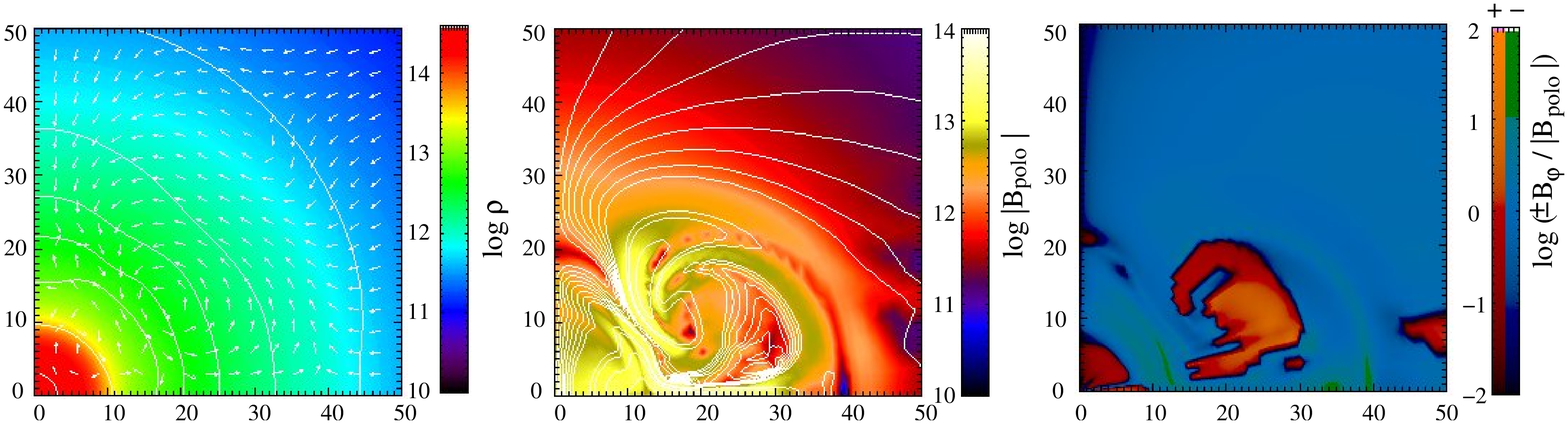}}
  \caption{Configuration of the innermost region of the collapsed star
    at the end of the evolution for models A1B3G5-D3M0 (top panels;
    $ t = 60 \mathrm{\ ms} $) and s20A1B1-D3M0 (bottom panels;
    $ t = 142.5 \mathrm{\ ms} $). The left panels show the rest mass
    density as $ \log \rho $ in units of $ \mathrm{g\ cm}^{-3} $,
    overplotted by the meridional velocity field $ (v^r, v^\theta) $
    (arrows), and isocontours of the specific internal energy
    $ \epsilon $. The center panels display the logarithm of the
    poloidal component of the magnetic field,
    $ \log{|B_\mathrm{polo}|} $, in units of G and the magnetic field
    lines in the $ r $--$ \theta $ plane. The right panels show
    $ B^{\varphi} / |B_\mathrm{polo}| $. All axes are in units of km.}
  \label{fig:morphology}
\end{figure*}

The main qualitative differences between the various models become
apparent when we study the detailed structure of the magnetic field of
the resulting PNS. In Fig.~\ref{fig:morphology} we show
two-dimensional snapshots of selected hydrodynamic and magnetic field
variables at the final time of the simulations for two representative
models of our sample, namely model A1B3G5-D3M0 (top panels) and model
s20A1B1-D3M0 (bottom panels). For typical simulations with initial
poloidal magnetic fields (D3M0 models) the resulting PNS has two
clearly distinct parts (see left panels of Fig.~\ref{fig:morphology}):
an inner region with a size of $ \sim 10 \mathrm{\ km} $, where
nuclear density is exceeded and which is almost rigidly rotating, and
a surrounding shell extending to the neutrino sphere at
$ \sim 30 \mathrm{\ km} $, with subnuclear densities and which is
strongly differentially rotating. These two parts are also visible in
the distribution of the magnetic field (see center and right panels of
Fig.~\ref{fig:morphology}). The inner region has a mixed toroidal and
poloidal magnetic field configuration, with both components having
similar strength, which results in a helicoidal structure aligned with
the rotation axis. As this part of the PNS is almost rigidly rotating
and practically in equilibrium, the magnetic field hardly evolves in
time. On the other hand, the outer shell is differentially rotating;
thus the toroidal magnetic field component dominates and grows
linearly with time due to the $ \Omega $-dynamo mechanism.

If we compare the microphysical with the simplified simulations, we
find that some significant morphological differences arise due to the
stronger convection in the microphysical models just below the
neutrino sphere. These motions affect the magnetic field, since they
twist the poloidal magnetic field lines, generating a much more
complicated structure of the poloidal field for those models. In
particular those strong meridional currents distort the magnetic field
in such a way that in some regions the poloidal component changes
direction with respect to the rotation axis (see e.g.\ bottom-right
panel of Fig.~\ref{fig:morphology}). This produces a negative effect
in the $ \Omega $-dynamo as in these regions the magnetic field is
wound up in the opposite direction. However, the overall
$ \Omega $-dynamo mechanism seems not to be affected in a significant
way by these local effects. 

Model A4B5G5-D3M0 has to be discussed separately, since it has
initially significantly stronger differential rotation and more
angular momentum than the other models. As a result this model
undergoes a core bounce due to centrifugal hang-up before reaching
nuclear density. Its structure is toroidal with an off-center
maximum density. Although it exhibits stronger differential rotation
at the beginning compared to the other models, and the amplification
process during collapse is thus more efficient, after bounce its
angular velocity $ \Omega $ is smaller (as the PNS is less compact) and
therefore the linear amplification due to $ \Omega $-dynamo is less
pronounced. The main differences in the magnetic field structure of
its PNS with respect to the other models are that, first, the
$ \Omega $-dynamo is active not only in the high-density torus, but
also in the central lower-density region, and, second, the strong
meridional currents twist the magnetic field lines around the
torus. However, we point out that in the investigated range of initial
rotation configurations all microphysical models are significantly less
influenced by rotation than the simplified models (like A4B5G5), and
that even for rather extreme rotation such collapse dynamics,
leading to a toroidal structure, is strongly suppressed if an advanced
description of microphysics is used \citep[which is in accordance with
the comprehensive parameter study by][]{dimmelmeier_07_a}.

In the models with initially purely toroidal field at the beginning
(T3M0 series), a poloidal field cannot emerge in axisymmetry. Hence,
the final magnetic field structure of the PNS consists of a stationary
and entirely toroidal magnetic configuration with the highest field
strengths found in the high density regions. As the rotational profile
does not affect the distribution of the magnetic field, the different
regions of the PNS are not visible in the structure of the magnetic
field.


\subsection{Comparison with Newtonian results}

In order to study the general relativistic effects in the evolution of
the magnetic field, we choose a subset of our simulations with the
hybrid EOS to represent the relativistic version of some of the models
of \citet{obergaulinger_06_b, obergaulinger_06_a}. Their
first paper is devoted to Newtonian simulations of magneto-rotational
core collapse, while in their second paper an effective relativistic
gravitational potential was used to mimic general relativistic effects
(while still keeping a Newtonian framework for the hydrodynamics; TOV
models in their notation). Since in contrast to their work we use the
passive field approximation, the comparison can only be made with the
low magnetic field models presented in that work, namely the ``M10''
models. In these models the magnetic field does not affect the
collapse dynamics and our approximation is valid. Although there are
no qualitative differences between Newtonian and general relativistic
models \citep[aside from those coming purely from the hydrodynamics as
described in][]{dimmelmeier_02_a, dimmelmeier_02_b}, some
dissimilarities can be found in the magnetic field strength and
amplification rates after core bounce.

\begin{figure*}[ht!]
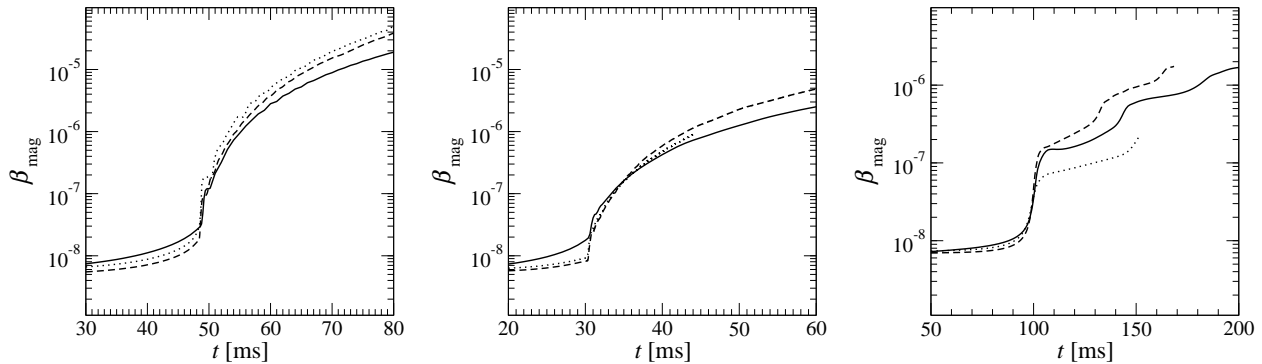

  \centering
  \resizebox{0.3\textwidth}{!}{\includegraphics*{7432f7a.eps}}~
  \resizebox{0.3\textwidth}{!}{\includegraphics*{7432f7b.eps}}~
  \resizebox{0.3\textwidth}{!}{\includegraphics*{7432f7c.eps}}
  \caption{Comparison of the time evolution of the magnetic energy
    $ E_\mathrm{mag} $ for models A1B3G3-D3M0 (left panel), A3B3G5-D3M0
    (center panel) and A2B4G1-D3M0 (right panel). The line styles
    represent simulations performed in general relativity (solid
    lines), a purely Newtonian treatment (dotted lines), and Newtonian
    hydrodynamics with an effective relativistic TOV potential (dashed
    lines).}
  \label{fig:betamag_comp}
\end{figure*}

We have studied the evolution of the magnetic energy parameter
$ \beta_\mathrm{mag} $ for the various models, and plot the results in
Fig.~\ref{fig:betamag_comp}. Note that for the same initial magnetic
field, the magnetic field contribution to the magnetic energy parameter
differs between a purely Newtonian treatment, a Newtonian formulation
with the effective relativistic TOV potential, and general relativity.
As a consequence, the initial value of $ \beta_\mathrm{mag} $ is
not the same in these three cases. In order to be able to make an
unambiguous comparison, we scale the magnetic fields such that
$ \beta_\mathrm{mag} $ in the initial model is equal to the value in
general relativity. In general, for a similar hydrodynamic behavior
(models A1B3G3-D3M0 and A3B3G5-D3M0) the magnetic energy attained
during the evolution is smaller in the general relativistic case than
in the Newtonian case (with either regular or effective relativistic
TOV potential).

The winding up of magnetic field lines is the main mechanism
responsible for the increase of the magnetic field during the
collapse. Therefore the amplification rate for
$ \beta_\mathrm{mag} $ is determined by what rotation rate is reached
and also by how strongly the poloidal component of the magnetic field
is compressed. In the general relativistic case both higher densities
and also stronger rotation are achieved \citep{dimmelmeier_02_b}.
To investigate the impact of general relativistic gravity on the
magnetic field compression, we consider $ \beta_\mathrm{polo} $ as
this quantity is the seed for the $ \Omega $-dynamo. In general
relativity the PNS has in general a smaller mass $ M_\mathrm{PNS} $
than in the corresponding Newtonian simulation of the same
model. Following the relation established in Sect.~\ref{sec:beta_evol}
(see the bottom panel of Fig.~\ref{fig:beta_mcollapse_comp2}), the
smaller PNS mass in the general relativistic simulation leads to a
lower value of $ \beta_\mathrm{polo} $. Therefore in that case,
despite the larger $ \Omega_\mathrm{c} $ the much smaller magnitude of
$ \beta_\mathrm{polo} $ results in a longer time scale for the
$ \Omega $-dynamo via Eq.~(\ref{upper_limit}), and hence a smaller
growth rate of the magnetic field.

In the multiple centrifugal bounce model A2B4G1-D3M0, general
relativistic effects lead to a bounce at significantly higher maximum
densities than in Newtonian gravity. Therefore, this is the only
investigated model where $ M_\mathrm{PNS} $, and consequently
$ \beta_\mathrm{polo} $ as well as $ \beta_\mathrm{mag} $ are larger
in the general relativistic simulation.


\subsection{Gravitational waves}
\label{gravitational_waves}

\begin{figure}[t!]
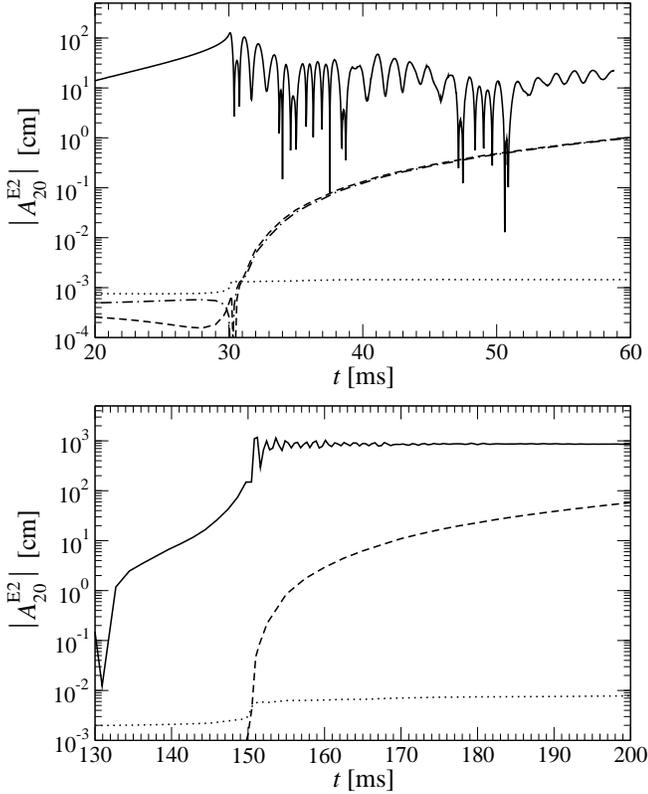

  \centering
  \resizebox{0.475\textwidth}{!}{\includegraphics*{7432f8a.eps}} \\ [0.5 em]
  \resizebox{0.475\textwidth}{!}{\includegraphics*{7432f8b.eps}}
  \caption{Absolute value of the gravitational wave amplitude
    $ A^\mathrm{E2}_{20} $ (solid line) for models
    A1B3G5-D3M0/T3M0/DT3M0 (top panel) and s20A1B5-D3M0/T3M0 (bottom
    panel). For the low magnetic field strengths considered, the
    contribution of the magnetic field to the waveform is negligible,
    and the signals of series D3M0, T3M0, and DT3M0 are practically
    identical to the purely hydrodynamic waveform. For clarity, the
    component $ A^\mathrm{E2}_\mathrm{20\,mag} $ from the magnetic
    field is also plotted for models A1B3G5-D3M0 (top panel, dashed
    line), A1B3G5-T3M0 (top panel, dotted line), A1B3G5-DT3M0 (top
    panel, dashed-dotted line), s20A1B5-D3M0 (bottom panel, dashed
    line), and s20A1B5-T3M0 (bottom panel, dotted line).}
  \label{fig:mcollapse_gw_1}
\end{figure}

We calculate the gravitational wave output from all of our simulations
using the Newtonian quadrupole formula given in
Eq.~(\ref{eq:mag_stress_formula}), which includes the magnetic terms.
Thus, the quadrupole wave amplitude $ A^\mathrm{E2}_{20} $,
which is related to the dimensionless quadrupolar strain amplitude
$ h^\mathrm{quad} $ as
\begin{equation}
  h^\mathrm{quad} = \frac{1}{8} \sqrt{\frac{15}{\pi}}
  \sin^2 \theta \frac{A^\mathrm{E2}_{20}}{R},
\end{equation}
contains the contribution $ A^\mathrm{E2}_\mathrm{20\,mag} $
corresponding to the magnetic field. Here $ h^\mathrm{quad} $ is the
only independent component of the radiative part
$ h^\mathrm{quad}_{ij} $ of the spatial metric as given by
Eq.~(\ref{eq:quad_formula}). In order to understand how the magnetic
field affects the waveforms, we also separately compute
$ A^\mathrm{E2}_\mathrm{20\,mag} $. The resulting waveforms for some
representative models are shown in Fig.~\ref{fig:mcollapse_gw_1}. As
the magnetic field is very low at all times, $ b^2 \ll \rho $, the
component of the gravitational wave due to the magnetic field is
several orders of magnitude smaller than $ A^\mathrm{E2}_{20} $.

Therefore, during the core bounce and the immediate post-bounce phase,
the waveforms we obtain are practically identical to the ones
presented for the same model setup in \citet{dimmelmeier_02_a} (for
the simplified models) and \citet{dimmelmeier_07_a} (for the
microphysical models), which can also be downloaded from a
freely accessible waveform catalog at
\texttt{www.mpa-garching.mpg.de/rel\_hydro/}\linebreak[1]%
\texttt{wave\_catalog.shtml}. The values for
$ A^\mathrm{E2}_{20} $ lie in the range between about
$ 30 \mathrm{\ cm} $ and $ 3000 \mathrm{\ cm} $, which translates to a
$ h^\mathrm{quad} $ of roughly $ 3 \times 10^{-22} $ to
$ 3 \times 10^{-20} $ (assuming a distance $ R = 10 \mathrm{\ kpc} $
to the source and optimal orientation between the source and the
detector).

We also emphasize that all investigated microphysical models yield
gravitational wave signals known as Type~I in the literature, i.e.\
the waveform exhibits a positive pre-bounce rise and then a large
negative peak, followed by a ring-down. This is to be expected, as
recent studies using the same hydrodynamical model setup
\citep{ott06a, dimmelmeier_07_a} have shown that the inclusion of
microphysics in stellar core collapse simulations suppresses the other
signal types, which were associated to multiple centrifugal bounce
(Type~II signals) or rapid collapse with a very small mass of the
inner core (Type~III signals).

After bounce, the star reaches a quasi-equilibrium state, and thus,
the hydrodynamic component of the waveform decreases. At the same
time, for models D3M0, the magnetic field grows linearly with
time. Such a behavior in the magnetic field produces an increasing
gravitational wave signal, which grows quadratically with time due to
the dependence on the magnetic field in Eq.~(\ref{eq:mag_stress_formula}). 
However, at the end of the simulation, the magnetic field component of
the waveform is still negligible in comparison with the hydrodynamic
component. It is expected that at later times, as the amplification of
the magnetic field reaches saturation, the influence of the magnetic
field on the waveform becomes significant, both due to its effect on
the dynamics and also due to the contribution of the magnetic field to
the gravitational radiation itself. We note, however, that the effect
of the MRI could additionally lead to noticeable changes in the
waveforms, provided it were able to efficiently amplify the magnetic
field (see discussion in Sect.~\ref{subsec:mri}).

For models T3M0, on the other hand, the component of the waveform due
to the magnetic field is even smaller than for the D3M0 models. This
is a consequence of the inefficient amplification of the magnetic
field via the radial compression. After bounce, the magnetic component
of the waveforms in models T3M0 does not grow, and hence it is not
expected to dominate the waveform later in the evolution, unless other
processes amplifying the magnetic field were present.


\subsection{Amplification of the magnetic field}
\label{sub:ampli}

Different mechanisms that amplify the magnetic field can act during a
core collapse or the subsequent evolution of the newly formed
PNS. This issue is of great importance, since the evolution of the PNS
during its first minute of life until a cold NS forms can change
drastically depending on the initial conditions at formation. One of
the most important aspects is the distribution of angular momentum. A
highly differentially rotating PNS can be subject to various types of
instabilities, such as the dynamical low-$ \beta $ instability, the
classical bar-mode instability (which is unlikely to occur in a PNS on
dynamical time scales as it requires very high values of $ \beta $),
or the secular CFS instability. Such instabilities are potential
sources of detectable gravitational radiation. Therefore, a natural
question that arises is whether the magnetic field is going to grow
sufficiently fast to act on the PNS dynamics by flattening the
rotation profiles (and therefore preventing the instabilities to
develop), or whether, instead, the growth process may take a few
seconds, allowing the instabilities to grow and the accompanying
gravitational waves to become detectable. A number of effects can
amplify the magnetic field shortly after PNS formation. In the
following, we discuss these effects and estimate their importance for
our models of core collapse\footnote{For these estimates we utilize
  the Newtonian limit, since most of the work on linear analysis of
  instabilities has not yet been extended to general relativity.
  Furthermore, for an approximate assessment, the restriction to a
  Newtonian treatment appears sufficiently accurate.}.


\subsubsection{\boldmath $ \Omega $-dynamo}

Within our passive field approximation we can only compute the
amplification rates for the $ \Omega $-dynamo, for which the magnetic
field grows linearly with time; therefore $ \beta_\mathrm{mag} $
grows quadratically with time (see Appendix~\ref{app:odynamo}). The
time scale $ \tau_\Omega $ of this amplification process and the
estimated time $ t_\mathrm{sat} $ at which the field saturation begins 
are given in Table~\ref{tab:MCC_models}. In the fastest case of our
model sample, which occurs for model A1B3G3-D3M0, saturation is
reached at about $ 300 \mathrm{\ ms} $, and in most other cases, the
$ \Omega $-dynamo saturates at times larger than $ 0.5 \mathrm{\ s} $.
Note, however, that these estimates depend on the initial magnetic
field strength, which is chosen to be
$ B^*_0 = 10^{10} \mathrm{\ G} $. For lower values of the
magnetic field these time scales can be scaled as
(see Eq.~\ref{eq:tau_omega})
\begin{equation}
  \tau_{\Omega} \approx
  \tau_{\Omega \, 10}
  \left( \frac{10^{10} \mathrm{\ G}}{B^*_0} \right),
  \qquad
  t_\mathrm{sat} \approx t_\mathrm{sat \, 10}
  \left( \frac{10^{10} \mathrm{\ G}}{B^*_0} \right).
\end{equation}
We recall that stellar evolution calculations predict that in a
progenitor core the poloidal component of the magnetic field can
initially have a strength of about $ 10^6 \mathrm{\ G} $
\citep{heger05}. For such an initial magnetic field the saturation
time scale becomes several hours. This makes the $ \Omega $-dynamo a
very inefficient mechanism to amplify the magnetic field during core
collapse and bounce, unless the progenitors are highly magnetized
($ B > 10^{10} \mathrm{\ G} $) for which the saturation could be
reached within a few dynamical time scales. The magnetic field at the
saturation is independent of the initial magnetic field strength and
of the order of $ \sim 10^{16} \mathrm{\ G} $.


\subsubsection{Magneto-rotational instability}
\label{subsec:mri}

There are other magnetic field amplification processes that our
simulations cannot account for, but for which it is nevertheless
possible to estimate the growth rates. It has been suggested that the
magneto-rotational instability could amplify the magnetic field from
arbitrary weak fields up to values where equipartition between the
magnetic field energy and the rotational kinetic energy is reached
\citep{akiyama03}. However, our analysis shows that in the context of
core collapse such an amplification is still an open issue. We proceed
next to describe the MRI and the uncertainties related to its effect
on the amplification of the magnetic field in core collapse.

\paragraph{Linear regime:}

The MRI is a shear instability that generates turbulence and results
in an amplification of the magnetic field in a differentially rotating
magnetized plasma \citep{balbus91, balbus92}, redistributing angular
momentum in the plasma. Linear analysis shows that if the magnetic field
strength is very low, as in our case, the stability criteria for the 
MRI in the Newtonian limit \citep{balbus95} are
\begin{equation}
  \begin{array}{rcl}
    \mathcal{C}_\mathrm{MRI1} & = & \mb{g} \cdot {\bf \mathcal{B}} +
    \mathcal{R} \cdot \nabla\varpi > 0,
    \\ [0.2 em]
    \mathcal{C}_\mathrm{MRI2} & = & (\mb{g} \times \nabla \varpi) 
    (\mathcal{B} \times \mathcal{R}) > 0,
  \end{array}
  \label{eq:mricriteria}
\end{equation}
where $ \mathcal{R}= \varpi \nabla (\Omega^2) $. Note that these
criteria are very similar to the Solberg--H{\o}iland
criteria~(\ref{eq:shcriteria}) for convection, but use an angular
velocity gradient $ \mathcal{R} $ instead of an angular momentum
gradient $ \mathcal{J} $. Since in the core collapse scenario
$ \mathcal{R} \le 0 $ is satisfied almost everywhere, it is important
to compute the buoyancy terms given by $ \mathcal{B} $ to estimate the
onset of the MRI. For regions with $ \mathcal{B} > 0 $ (i.e\ with
a negative entropy gradient that is strong enough to compensate the
positive electron fraction gradient term in
Eq.~(\ref{eq:buoyancy_reformulation})), the first criterion is not
fulfilled. Furthermore, for regions with $ \mathcal{B} < 0 $ (i.e.\ a
positive or sufficiently small negative entropy gradient), the second
criterion is neither satisfied. This means that the presence of a
adequately strong negative entropy gradient (which also leads to
convective instability) enhances the MRI, although a positive entropy
gradient does not affect the condition for MRI instability. Note that
this peculiarity is caused by the negative value of $ \mathcal{R} $,
and does not happen in the Solberg--H{\o}iland
criteria~(\ref{eq:shcriteria}) for convection, as $ \mathcal{J} > 0 $
in that case. If at least one of the criteria~(\ref{eq:mricriteria}) is not satisfied
\emph{and} a magnetic field is present, then fluid and magnetic field
perturbations grow exponentially in time. Neglecting buoyancy terms,
the time scale for the fastest growing unstable mode can be roughly
estimated as\footnote{We note that~\citet{balbus91} derived a
  complicated expression including bouyancy terms which, however, is
  only valid in the equatorial plane. To the best of our knowledge the
  timescale for the fastest growing mode in the general case has not
  been computed so far. It would require solving the dispersion
  relation, a task out of the scope of this paper.}
\begin{equation}
  \tau_\mathrm{MRI} = 4 \pi \,
  \left| \frac{\partial\Omega}{\partial\ln{\varpi}} \right|^{-1},
\end{equation}
which is independent of the magnetic field configuration and strength.
Only those modes with a length scale larger than a critical wavelength
will grow \citep{balbus91}. This length scale can roughly be estimated
as $ \lambda_\mathrm{MRI} \sim 2 \pi c_\mathrm{A} / \Omega $, where
$ c_\mathrm{A} = \sqrt{B^2 / \rho} $ is the Alfv\'en speed. For the
typical values attained in the nascent PNS, in which the dominant
magnetic field is toroidal, the critical length scale is
\begin{equation}
  \lambda_\mathrm{MRI} \approx
  62 \left( \frac{B^{*\,0}}{10^{10} \mathrm{\ G}} \right)
  \left( \frac{1 \mathrm{\ ms}^{-1}}{\Omega} \right)
  \left( \frac{10^{14} \mathrm{\ g\ cm}^{-3}}{\rho} \right)^{1/2}
  \mathrm{\ m}.
\end{equation}
Note that we have scaled the length scale with the typical magnetic
field strength $ B^{*\,0} $ of the progenitor, and not with that of
the PNS itself. For the poloidal component and realistic values of the
initial magnetic field ($ B^{*\,0} \sim 10^6 \mathrm{\ G} $) this
length scale is reduced by several orders of magnitude
($ \lambda_\mathrm{MRI\,polo} \sim 0.6 \mathrm{\ cm} $). In any case,
resolving the scales needed to simulate the MRI is a challenging
problem as, in the case of weak magnetic fields, the wavelength of the
fastest growing mode (which is close to the critical length scale) is
typically much smaller than the available grid resolution.

\paragraph{Non-linear regime:}

Linear analysis provides tools to determine the onset of the
instability and the typical time and length scales. However, once the
perturbations of the magnetic field reach values comparable to the
magnetic field itself, linear analysis is no longer valid (although
in the weak field case the perturbations of the fluid variables are
still small). The amplification of the magnetic field due to the MRI
is therefore a nonlinear effect, and can only be studied by means of
numerical simulations. The appropriate numerical approach, due to the
smallness of the length scales necessary to be resolved, are local
simulations of the MRI in a shearing box. Numerical simulations of
this kind in three dimensions have been performed by
\citet{Hawley1995} in the context of accretion discs. They show that
if the instability condition of linear analysis is fulfilled, then the
amplification of the magnetic field proceeds by the formation of an
axisymmetric channel flow. This is well understood, since the linear
MRI solution is also a solution of the nonlinear axisymmetric MHD
equations \citep{Goodman1994}. In the ideal MHD limit, the
amplification saturates when nonaxisymmetric perturbations destroy the
channel flow. It is important to emphasize the necessity of performing
three-dimensional simulations in the shearing box since, in
axisymmetry, the channel flow is not destroyed and any magnetic field
is able to grow continuously, reaching saturation only when the MRI
length scale is of the order of the region in which the MRI is present
\citep{Hawley1992}.

\begin{figure*}[t!]
  \centering
  \resizebox{0.475\textwidth}{!}{\includegraphics*{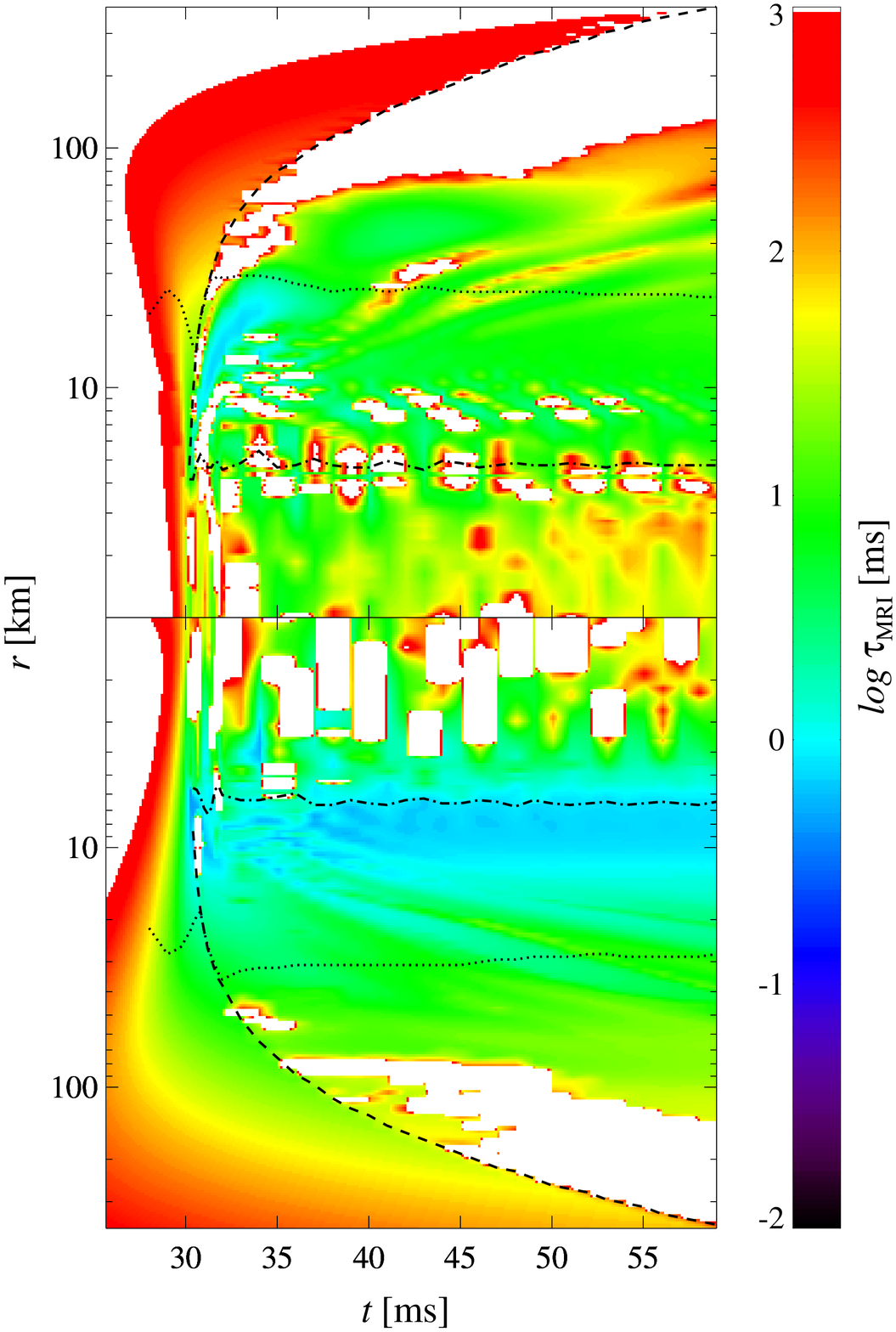}}
  \resizebox{0.475\textwidth}{!}{\includegraphics*{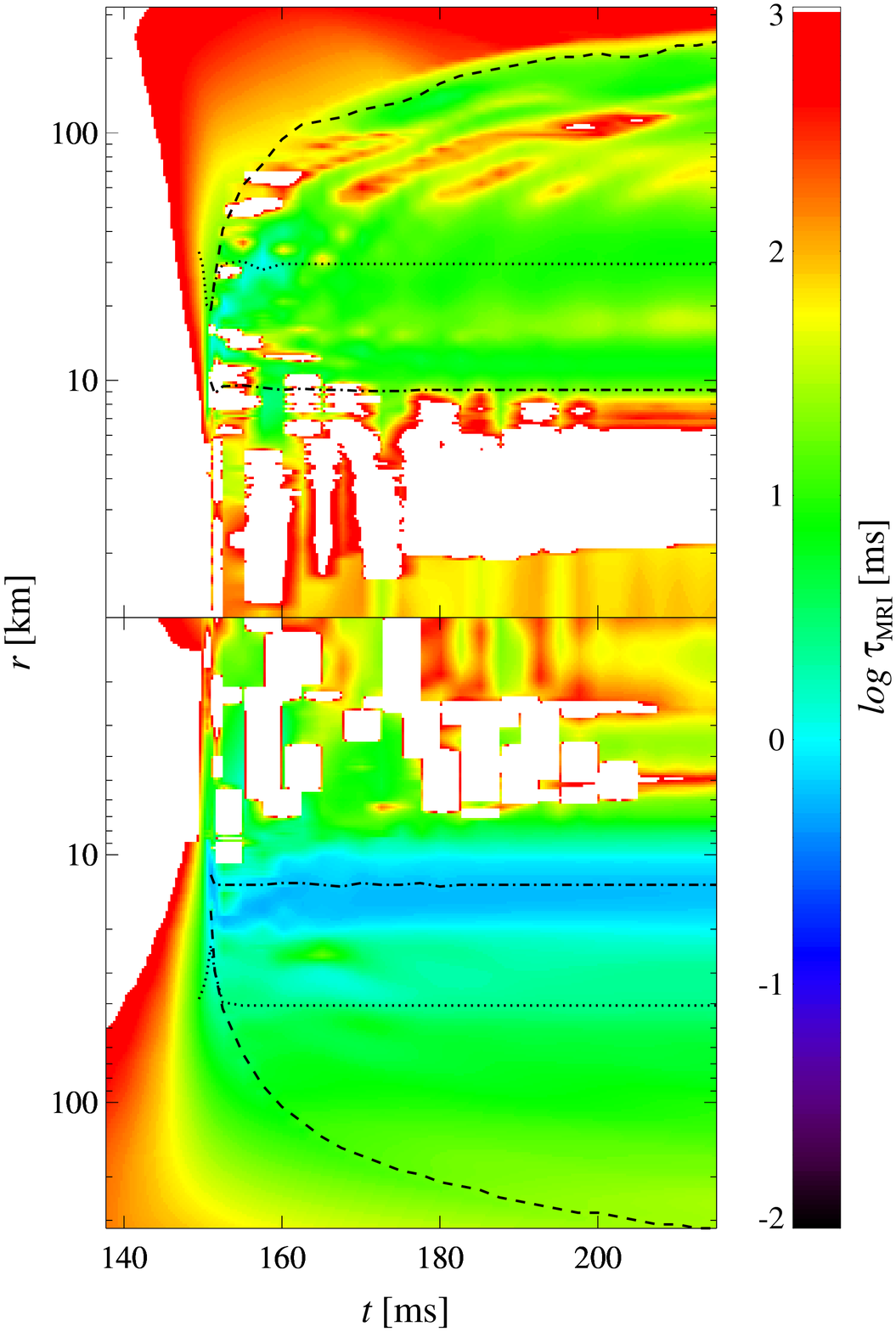}}
  \caption{Time evolution of the angular averaged value of the growth
    time scale $ \tau_\mathrm{MRI} $ of the fastest growing mode of
    the MRI for models A1B3G5-D3M0 (left panel) and s20A1B5 (right
    panel). White regions are stable to the MRI. The top half panels
    show angular averages of $ \tau_\mathrm{MRI} $ near the pole
    ($ 0 < \theta < \pi / 6 $), while the bottom half panels show
    these averages near the equator ($ \pi / 3 < \theta < \pi / 2 $).
    The shock radius (dashed lines), neutrino sphere radius (dotted
    line), and radius of shock formation (dash-dotted line) are also
    displayed.}
  \label{fig:MRI}
\end{figure*}

For a magnetic field distribution with zero mean at large scales, the
amplification proceeds from arbitrarily weak fields \citep{Hawley1996}
and saturates irrespective of the initial magnetic field at average
values of $ P_\mathrm{mag} / P_0 \sim 0.01 $, where $ P_0 $ is the
initial gas pressure. If a mean magnetic field is present (as in our
case), the saturated magnetic field depends on the initial magnetic field
strength. In the most favorable case of a vertical magnetic field, the
total amplification by the MRI is only about a factor 20 of the
original field, this amplification being even smaller in the case of a
purely toroidal field \citep{Hawley1995}. On the other hand,
\citet{Sano2004} have suggested that for sufficiently weak magnetic
fields the saturation level could be independent of the initial field
and equal to that in the zero-mean case. If this were confirmed it
would mean that, for the weak magnetic field strength present in our
magneto-rotational core collapse models
($ P_\mathrm{mag} / P_0 \sim 10^{-8} $ in the PNS for progenitors with
$ B^{*\,0} = 10^{10} \mathrm{\ G} $), a magnetic field of
$ \sim 10^{16} \mathrm{\ G} $ could be reached on time scales of
$ \tau_\mathrm{MRI} $. Such a strong magnetic field would have a
significant effect on the dynamics, similar to that observed in
numerical simulations with highly magnetized progenitors
\citep{obergaulinger_06_a,obergaulinger_06_b,shibata_06_a}. In the
opposite case, the MRI would fail to considerably amplify the magnetic
field, and for a purely toroidal field the magnetic field should grow
only by a factor of about 3 according to \citet{Hawley1995}.

The inclusion of more complex physics relevant for the core collapse
scenario (like radiation, diffusion, or resistivity) can significantly
change the amplification process, since in the nonideal case the
reconnection of magnetic field lines seems to be the dominant effect
in the saturation process of the MRI. In general, these effects act
towards lowering the values of the saturation; for reasons of
simplicity we do not consider them in this discussion \citep[see][and
references therein for a detailed review on this topic]{Hawley2005}.

Furthermore, it has to be noted that all local simulations of the
MRI have been performed in the context of Keplerian accretion discs,
and, hence, some of the underlying physical conditions are not valid
in the case of a PNS. For example, the typical sound speed
$ c_\mathrm{s} $ in those simulations is of the order of $ 10^{-3} $.
Only the parametric study performed by \citet{Sano2004} covers a
wider range of values of
$ c_\mathrm{s} \sim 10^{-8} \mbox{\,--\,} 10^{-2} $ in units of $ c $.
However, the sound speed in a PNS is higher,
$ c_\mathrm{s} \sim 10^{-1} $. Rotational velocities and profiles are
also very different in a disc and a PNS. Therefore, appropriate
local simulations of the PNS scenario should eventually be performed
in order to confirm the growth of the MRI for a weakly magnetized PNS,
and to infer the magnetic field at which the instability saturates.

\paragraph{Our results:}

As the MRI involves a backreaction of the magnetic field onto the
dynamics, we cannot study this effect in our simulations, as we assume
the passive field approximation. Furthermore, even with ``active''
magnetic fields, both the resolution needed to resolve the MRI length
scale ($ \sim 10 \mathrm{\ m} $) and the requirement for
three-dimensional simulations are not affordable with present
computers. Therefore, we are limited to analyzing whether our
magnetized collapse models are susceptible to developing such an
instability according to linear analysis estimates, leaving aside the
issue of saturation, whose uncertainties need a deeper analysis which
is beyond the scope of this work. In order to estimate how MRI could
change our results if it were taken into account properly, we
determine the regions where the MRI instability
criteria~(\ref{eq:mricriteria}) are not satisfied. Inside these
regions we calculate the time scale for the fastest growing mode. In
Fig.~\ref{fig:MRI} we show the results for the models A1B3G5-D3M0
(left panel) and s20A1B5-D3M0 (right panel). We note that since the
onset of the MRI is independent of the magnetic field strength,
provided that a poloidal component exists, any composition of D3M0 and
T3M0 models has the same instability properties as the D3M0 models.

Our analysis of all computed models shows that during the infall phase
the MRI is either not possible or the typical time scales involved are
much larger (i.e.\ $ > 10 \mathrm{\ s} $) than the duration of the
collapse itself. Therefore the instability can affect neither the
dynamics nor the magnetic field strength in that phase. Around the
time of core bounce, the angular velocity gradient is larger and the
MRI time scale becomes dynamical. Almost the entire region between the
shock formation radius (at $ \sim 10 \mathrm{\ km} $) and the shock
itself is MRI unstable with time scales of the order of
$ \sim 1\mbox{\,--\,}10 \mathrm{\ ms} $. Note that the innermost part
of the PNS rotates rigidly, and therefore the MRI unstable region that
appears in the inner $ 2 \mathrm{\ km} $ is possibly a numerical
artifact caused by the probably unphysical negative entropy gradient
mentioned in Sect.~\ref{subsec:convection}. Some differences appear
when comparing microphysical and simplified models.

A general feature of the microphysical models is the post-bounce
appearance of a negative entropy gradient (regions with $ N^2 < 0 $,
see Sect.~\ref{subsec:convection}). This property is much less
prominent in the simplified models (except in model A1B3G5). Thus, the
presence of a such gradient in the microphysical models enhances the
occurence of the MRI behind the shock as compared to the simplified
models (see Fig.~\ref{fig:MRI}), since in these regions the cause for
the instability is mainly the presence of a negative entropy
gradient. Around the neutrino sphere the presence or absence of
a negative entropy gradient does not affect the onset of the
instability since it is caused by the strong negative
angular velocity gradient. Therefore, only small differences can be
found in the latter region between the simplified and the
microphysical models.

As a result of this analysis, for collapse progenitors with a magnetic
field smaller than $ 10^{10} \mathrm{\ G} $ (hence including
astrophysically more relevant initial values of $ 10^6 \mathrm{\ G} $),
we infer that perturbations of the magnetic field are going to grow
exponentially on dynamical time scales and will reach saturation in the
unstable regions mentioned above. However, the value of the magnetic
field at which saturation appears is still unknown, which is a key
issue in order to establish the effects of the MRI, if any, on the
dynamics. Nevertheless, even if the MRI were unable to considerably
amplify the magnetic field, it could still play a major role at late
times during the evolution of the PNS, provided other amplification
mechanisms were capable to increase the magnetic field to significant
larger values (see below). In such a situation the MRI could have an
impact on the dynamics by transporting angular momentum outwards and
driving the PNS towards rigid rotation.


\subsubsection{Dynamo mechanisms}

The wind-up process of the magnetic field ($ \Omega $-dynamo)
discussed before is a mechanism that works by transforming the
poloidal magnetic field into a toroidal field and extracting energy
from differential rotation. In axisymmetry this process amplifies the
magnetic field linearly with time as long as differential rotation
exists. If the axisymmetry condition is relaxed, however, a number of
instabilities of the toroidal field can transform the toroidal
magnetic field back into a poloidal magnetic field. This feedback then
``closes'' the dynamo process.

The first group of instabilities are those related to convective
unstable regions, neutron-finger instabilities (due to a negative
composition gradient) and, in general, turbulence. In these cases the
$ \alpha $-effect is the one which closes the dynamo in the
$ \alpha\mbox{-}\Omega $-dynamo \citep{Thompson1993}. Computations of
this effect \citep{bonanno05} suggest that even for a rapidly rotating
PNS with a period around $ 1 \mathrm{\ ms} $ (i.e.\ comparable to the
models presented here), the time scale for the growth of the magnetic
field is $ \sim 1 \mathrm{\ s} $. Therefore, this effect is probably
not important after core bounce on dynamical time scales. However, for
larger time scales (i.e.\ several seconds), and if the MRI is not
efficient enough, this mechanism will most likely amplify the magnetic
field, leading to magnetic braking of the PNS within a few seconds.

There are also types of instabilities that can act in stably
stratified regions, i.e.\ regions which are convectively stable.
\citet{spruit99} has proposed the Tayler
instability \citep{tayler73} as a mechanism to close the dynamo. This
dynamo has been confirmed in numerical simulations by
\citet{braithwaite06b, braithwaite06a}. The condition for this kink-type 
instability to grow in the rotating case ($ m = 1 $ mode) is
\citep{spruit99}
\begin{equation}
  \partial_\theta \ln B_\varphi^2 \, \sin \theta \, \cos \theta > 0,
\end{equation}
which is satisfied almost everywhere inside the star in our
simulations. The growth rate of the instability is of the order of
the Alfv\'en time scale,
\begin{equation}
  \tau_\mathrm{T} = \frac{2 \pi}{\Omega_\mathrm{A}}
  \quad
  (\Omega \ll \Omega_\mathrm{A}),
  \quad \quad
  \tau_\mathrm{T} = \frac{2 \pi \Omega}{\Omega_\mathrm{A}^2}
  \quad
  (\Omega \gg \Omega_\mathrm{A}),
  \label{eq:tayler_growth_time}
\end{equation}
where $ \Omega_\mathrm{A} = c_\mathrm{A} / R $ and $ R $ is the
typical size of the region considered. In case this instability
appears, it destroys the toroidal magnetic field by transforming it
into a poloidal field which feeds back the amplification of the
toroidal magnetic field via the $ \Omega $-dynamo. Therefore, the
dynamo is only effective if the $ \Omega $-dynamo is able to generate
a toroidal magnetic field faster than the Tayler instability destroys
that field, i.e.\ $ \tau_\mathrm{T} \gg \tau_\Omega $. Saturation is
then reached as $ \tau_\mathrm{T} \approx \tau_\Omega $. Note that
depending on the system, the saturated magnetic field can be weak
enough not to affect the dynamics.

If we consider the typical toroidal magnetic field at bounce to be
$ 10^{13} \mathrm{\ G} $ (as in the T3M0 models) with a typical
density in the PNS of
$ \rho \sim 2 \times 10^{14} \mathrm{\ g\ cm}^{-3} $ and a typical
size of the inner region of $ R \sim 10 \mathrm{\ km} $, then the time
scale for the growth of the Tayler instability is strongly increased
by rotation,
\begin{equation}
  \tau_\mathrm{T} \approx
  3 \left( \frac{10^{10} \mathrm{\ G}}{B^{*\,0}} \right)^2
  \left( \frac{R}{10 \mathrm{\ km}} \right)^2
  \left( \frac{\Omega_\mathrm{c}}{1 \mathrm{\ ms}^{-1}} \right)
  \mathrm{\ hr},
\end{equation}
which we obtain from the $ \Omega \gg \Omega_\mathrm{A} $ limit of
Eq.~(\ref{eq:tayler_growth_time}).

This means that for a typical progenitor with a toroidal magnetic
field of $ B^{*\,0}_\varphi \sim 10^{10} \mathrm{\ G} $, the
instability proposed by \cite{spruit99} is going to be very
inefficient in amplifying the magnetic field. However, on a longer
time scale, when other mechanisms could amplify the magnetic field
(e.g.\ the $ \alpha\mbox{-}\Omega $-dynamo), the Tayler instability
could also become important.


\section{Conclusions}
\label{sec:conclusions}

In this paper we have presented numerical simulations of the collapse
of rotating magnetized stellar cores in the CFC approximation of
general relativity, as well as tests assessing our numerical approach
for solving the ideal general relativistic magneto-hydrodynamics
(GRMHD) equations.

As initial models we have set up (either fully or nearly) stationary
configurations of weakly magnetized stars in general relativity, with
either toroidal or poloidal (or both) magnetic field components. We
have used the ``test'' passive field approximation for evolving these
initial models, for which the magnetic pressure in all cases
considered is several orders of magnitude smaller than the fluid
pressure.

We have performed tests to check the accuracy and convergence
properties for the GRMHD extension of our code. For magnetic field
quantities we have found an order of convergence above 1 in all of the
performed tests. These results are consistent with the second-order
accuracy (in space and time) of our numerical scheme, reduced to first
order only at shocks and local extrema. The errors in all of the cases
in which the theoretical solution is known are below 0.1\%, except at
shocks, which are correctly captured within only few numerical
cells thanks to the use of high-resolution shock-capturing schemes.

For the simulations of magnetized core collapse, we have considered
cases with magnetic fields which are initially either purely poloidal
(series D3M0), purely toroidal (series T3M0), or a combination of
both. The D3M0 models are a general relativistic extension
of a subset of the cases evolved in fully coupled MHD by
\citet{obergaulinger_06_b, obergaulinger_06_a}, who used a Newtonian
formulation (approximating general relativistic effects to some extent
in the latter work). One of our aims has been to compare the dynamics
and gravitational waveforms with their results. No qualitative
differences have been found in the models studied, while
quantitatively the strength of the magnetic field at bounce and after
the collapse are consistently smaller in general relativity.

We have also compared simulations of models with improved microphysics
(employing a tabulated non-zero temperature equation of state (EOS)
and an approximate but effective deleptonization scheme) with the
simple (though still widely used) analytic hybrid EOS. The results
show that the microphysical models (i) lead to a more complex
structure of the poloidal magnetic field due to convective motions
surrounding the inner region of the PNS, and (ii) exhibit a wind-up of
the magnetic field ($ \Omega $-dynamo) that is more efficient than in
the simplified models for comparable rotation rates, which is due to
the larger compression of the poloidal component during the collapse.

We have found a unified explanation for the magnetic energy of all
models, independent of the description of gravity (general relativity
or Newtonian) or the EOS, which relates the angular velocity and mass
of the PNS with its magnetic energy and the growth rate of the
magnetic field due to the $ \Omega $-dynamo. This relation shows that
higher rotation rates and masses of the PNS lead to stronger magnetic
fields. We have shown that it is not possible to mimic the conditions
of the microphysical simulations using a simplified EOS. Simplified
models with a mass of the homologously collapsing inner core during
contraction and a mass of the PNS after bounce similar to the
respective masses of the microphysical models (and identical initial
rotation profiles) will undergo multiple centrifugal bounces, a
behavior that has recently shown to be an artifact of the neglect of
microphysics \citep{dimmelmeier_07_a}.

Further differences appear in the appearance of convective motion in
the PNS and behind the shock. This convection is more active in
microphysical models than in simplified ones. In models with slow
rotation, strong convection in the PNS occurs as a transient and
disappears within a few ten ms after bounce. Evidently, this transient
is an artifact as it does not appear in simulations of similar models
with comparable microphysics but using Boltzmann neutrino transport
\citep{mueller_04_a} instead of our simple advection scheme for the
electron fraction after core bounce. In rapidly rotating models
convection is not entirely suppressed by rotation but develops and
persists on longer time scales, albeit at a lower intensity.

As we have adopted the passive field approximation and the investigated
magnetic fields are weak in all phases of the collapse, the waveforms
of the gravitational radiation emitted by all our models are
practically identical to the corresponding ones in a purely
hydrodynamical simulation \citep{ott06a, dimmelmeier_07_a}, with the
contribution due to magnetic fields being several orders of magnitude
smaller than the total signal amplitude. However, if the MRI could
become dominant for the dynamics in the post-bounce phase, in a fully
coupled GRMHD simulation we would expect a clear imprint of such an
instability on the signal waveform. As expected, for the microphysical
models we obtain gravitational wave signals exclusively of Type~I, as all
other waveform types (in particular the Type~II signals associated with
multiple centrifugal bounces) are suppressed if more realistic
microphysics is taken into account.

For an astrophysically expected strength of the magnetic field
\citep{heger05}, where the initial toroidal component is much larger
than the poloidal one, we have obtained a topology of the magnetic
field in the PNS that is purely toroidal due to the radial compression
of the initial toroidal component. In this case the time scale for the
$ \Omega $-dynamo is very long (several hours). For progenitors with
stronger poloidal magnetic fields, we have found that a core/shell
structure is formed. Inside the core, where nuclear density is
exceeded, a mixed configuration of a poloidal and a toroidal magnetic
field yields a helicoidal configuration of the field lines. In the
surrounding shell (which extends several $ 10 \mathrm{\ km} $) the
poloidal magnetic field lines are wound up due to differential
rotation ($ \Omega $-dynamo), and shortly after core bounce the
magnetic field is dominated by the toroidal component. The growth time
scale for the toroidal component due to this process is, in the best
case scenario, several $ 100 \mathrm{\ ms} $.

We have also estimated the growth times for several other
instabilities that could appear if the passive field approximation or
the restriction to axisymmetry are removed. Among these the MRI is
apparently the fastest growing instability, although it remains
unclear if it is going to amplify the magnetic field sufficiently
(from the initially weak field values) to affect the dynamics at
all. In addition, we have found that the inclusion of microphysics
could enhance the MRI, since the regions behind the shock exhibit
a negative entropy gradient, resulting in a growth time of
$ \sim 10 \mathrm{\ ms} $ for the MRI. However, the influence of our
simplified neutrino treatment or the effects of an alternative
microphysical EOS must still be investigated in detail.

In the event that the MRI were unable to sufficiently amplify the
magnetic field in the PNS (which is still an open issue), the main
amplification mechanism would probably be the
$ \alpha\mbox{-}\Omega $-dynamo, which can amplify the magnetic field
to values where the magnetic energy is in equipartition with the
rotational kinetic energy on a time scale of, at least, several
seconds. The study of this effect is well beyond the goals of the
work presented in this paper, since the required time scales are much
longer than those affordable with current numerical
magneto-hydrodynamical (MHD) codes. Moreover, the underlying physics
necessary to be included (like neutrino transport, diffusion,
radiation, and cooling) is far more complex. However, in the light of
the results presented here, in which astrophysically expected values
for the magnetic field have been adopted, we can speculate about the
following scenario. If the MRI is ineffective, after core bounce the
magnetic field does not grow significantly strong during one (or maybe
several) seconds, and therefore differential rotation generated in the
collapse could persist. This ``one-second-window'' would provide
sufficient time for several instabilities to develop in the PNS. Such
instabilities are promising sources of gravitational waves.

The restriction to the passive magnetic field approximation in
studying magneto-rotational core collapse of weakly magnetized
progenitors can be justified if the MRI is indeed inefficient,
since none of the other estimated mechanisms seem to be able to
amplify the magnetic field significantly on dynamical time scales.
Otherwise, an ``active'' magnetic field approach becomes
necessary. However, it has to be stressed that the use of active
magnetic fields alone for core collapse simulations will probably not
be sufficient to model all the effects amplifying the magnetic field,
since the numerical resolution needed to correctly describe them
(probably less than $ 10 \mathrm{\ m} $) is not affordable in
current numerical simulations. In addition most of the prospectively
relevant effects have to be investigated in three dimensions, which
makes the computational task even more challenging.


\begin{acknowledgements}
  This research has been supported by the Spanish Ministerio de
  Educaci\'{o}n y Ciencia (grant AYA2004-08067-C03-01), by the DFG
  (SFB/Transregio 7 and SFB 375), by the DAAD and IKY (IKYDA
  German--Greek research travel grant), and by a Marie Curie
  Intra-European Fellowship within the 6th European Community
  Framework Programme (IEF 040464). It is a pleasure to thank
  C.~D.~Ott, A.~Marek, and H.-T.~Janka for their contributions related
  to the improved microphysics, L.~Ant\'on for many useful discussions
  in setting up the numerical scheme for the magnetic field evolution,
  M.~Obergaulinger for his help on understanding the MRI and the
  Newtonian simulations data, and E.~M\"uller as well as
  N.~Stergioulas for useful comments.
\end{acknowledgements}


\bibliography{7432}


\appendix

\section{Code tests}
\label{app:tests}

Here we discuss several tests we have designed in order to check the
accuracy of our numerical code when solving the induction equation
with the numerical methods described in this paper \citep[see
also][]{nfnr}. The ``toroidal test'' is set up for assessing the
ability of the code to maintain various magnetic field configurations
in equilibrium (labelled TTA and TTB) and to correctly compute the
amplification of the toroidal magnetic field as it is wound up by a
rotating fluid (TTC). On the other hand, the ``poloidal test'' (PT) is
designed to check whether the code can correctly compute the
compression of the poloidal magnetic field in a spherical
collapse. Finally, the strong spherical explosion test checks that the
code is able to handle the presence of radial shocks. We refer the
interested reader to \citet{nfnr} for details on the setup of the
toroidal and poloidal tests as well as the diagnostics we use to
compute the errors and order of convergence of our numerical schemes.


\subsection{Toroidal tests}

\begin{figure}[t!]
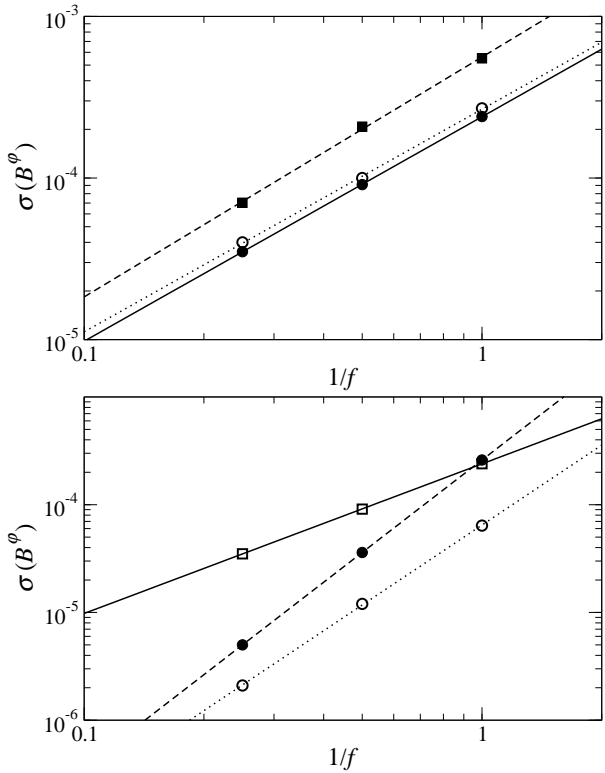

  \centering
  \resizebox{0.46\textwidth}{!}{\includegraphics*{7432f10a.eps}}
  \resizebox{0.46\textwidth}{!}{\includegraphics*{7432f10b.eps}}
  \caption{Global error $ \sigma $ in the toroidal magnetic field
    $ B^\varphi $ after a time evolution of $ 1 \mathrm{\ ms} $ as a
    function of $ 1 / f $ for a sequence of models with grid
    resolutions $ 80 \times 10 $ ($ f = 1 $), $ 160 \times 20 $
    ($ f = 2 $) and $ 320 \times 40 $ ($ f = 4 $). The top panel
    shows the error for the TTC test using different reconstruction
    schemes and the corresponding best fits to a power law: minmod
    (open circles, dotted line), MC (filled squares, dashed line), and
    PHM (filled circles, solid line). The bottom panel shows the error
    and respective fits using the PHM reconstruction for the TTA test
    (open circles, dotted line), TTB (filled circles, dashed line),
    and TTC (open squares, solid line).}
  \label{fig:TT}
\end{figure}

Fig.~\ref{fig:TT} shows the global error $ \sigma $ in the toroidal
magnetic field $ B^\varphi $ for the three tests TTA, TTB, and TTC
against $ 1 / f $ and the corresponding fits to a power law. Here
$ f $ denotes the factor which specifies the increase in resolution
from a coarse reference grid \citep[see][for details]{nfnr}. The
resulting convergence order of each numerical scheme (minmod, MC, and
PHM) as well as the errors for the highest resolution grid can be
found in Table~\ref{tab:TT}. Our results show that (i) the order of
convergence and the error is almost independent of the cell
reconstruction scheme employed, (ii) the order of convergence for the
TTC test is smaller than for the TTA and TTB tests, and (iii) the
order of convergence for the tests TTA and TTB is $ N > 2 $, and hence
higher than the theoretical expectation (which is second order, since
it is limited by the order of the time discretization, for which we
use a conservative, second order Runge--Kutta scheme).

\begin{table}[t!]
  \centering
  \caption{Convergence order $ N $ for the tests performed (TTA, TTB,
    TTC, and PT) and for different reconstruction procedures (minmod,
    MC, and PHM). The error $ \sigma_{320 \times 40} $ for the higher
    resolution grid is also given.}
  \begin{tabular}{cccccc}
    \hline \hline
    Test
    & Reconstruction scheme
    & $ N $
    & $\sigma_{320 \times 40} $ \\
    \hline
    TTA & minmod & 2.45 & $ 1.2 \times 10^{-6} $ \\
    TTA & MC     & 2.16 & $ 2.4 \times 10^{-6} $ \\
    TTA & PHM    & 2.46 & $ 2.1 \times 10^{-6} $ \\
    \hline
    TTB & minmod & 2.64 & $ 7.7 \times 10^{-6} $ \\
    TTB & MC     & 2.53 & $ 1.2 \times 10^{-5} $ \\
    TTC & PHM    & 2.85 & $ 0.5 \times 10^{-6} $ \\
    \hline
    TTC & minmod & 1.38 & $ 4.0 \times 10^{-5} $ \\
    TTC & MC     & 1.48 & $ 7.0 \times 10^{-5} $ \\
    TTC & PHM    & 1.39 & $ 3.5 \times 10^{-5} $ \\
    \hline
    PT  & minmod & 1.41 & $ 8.3 \times 10^{-4} $ \\
    PT  & MC     & 1.11 & $ 8.6 \times 10^{-4} $ \\
    PT  & PHM    & 1.17 & $ 8.6 \times 10^{-4} $ \\
    \hline \hline
  \end{tabular}
  \label{tab:TT}
\end{table}

\begin{figure}[t!]
  \centering
  \resizebox{0.49\textwidth}{!}{\includegraphics*{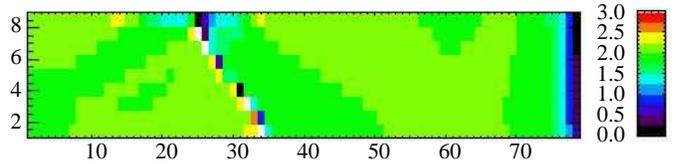}}
  \caption{Local order of convergence (color coded) for the TTA test
    after a total time evolution of $ 1 \mathrm{\ ms} $. White color
    is used for values $ \ge 3.0 $. The horizontal and vertical axes
    represent the number of cells of the reference grid in the radial
    and angular direction, respectively.}
  \label{fig:TTA_nij}
\end{figure}

The numbers reported in Table~\ref{tab:TT} demonstate that we obtain
similar results in all three tests for linear reconstruction schemes
(minmod and MC) and for the third order reconstruction scheme (PHM),
as the order of the scheme is limited by the second order
discretization in time and by the linear interpolation of the
cell-centered magnetic fluxes (which is a consequence of using a
staggered grid in the flux-CT scheme for the magnetic field). To
understand these results we note that in test TTC there is a component
of the magnetic field, $ B^{*\,\varphi} $, which grows linearly in
time, while in tests TTA and TTB no components evolve. Hence, the
order of convergence for the latter is higher than for test TTC. This
can be explained by investigating the \emph{local} order of
convergence, i.e.\ the order obtained when computing the error
$ \sigma_{ij} $ in each numerical cell instead of the global error
$ \sigma $. The results for test TTA are displayed in
Fig.~\ref{fig:TTA_nij} (similar plots can be obtained for the other
two cases). At some particular grid zones the order of convergence is
larger than two, while at most locations it remains around two.

\begin{figure}[t!]
  \centering
  \resizebox{0.47\textwidth}{!}{\includegraphics*{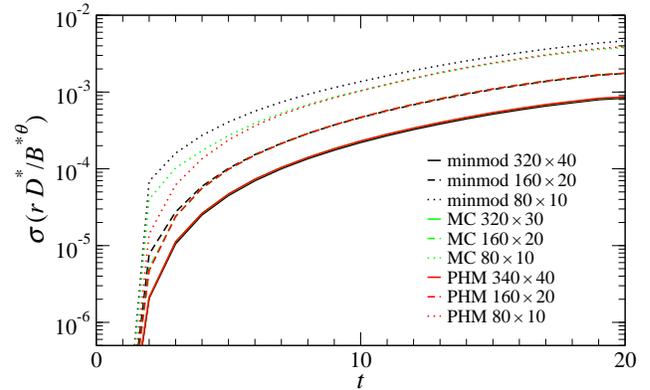}}
  \caption{Global error on $ r\, D^* / B^{*\,\theta} $ for the poloidal
    test (PT) as a function of time for three different grid
    resolutions: $ 80 \times 10 $, $ 160 \times 20 $, and
    $ 320 \times 40 $, and for four different reconstruction schemes
    (minmod, MC, and PHM).}
  \label{fig:PTA}
\end{figure}


\subsection{Poloidal test}

As mentioned before the setup and specifications of the poloidal test
are described in detail in \citet{nfnr}. Here we simply focus on
showing the comparison and performance of the various numerical
schemes employed in our simulations. \citep [Note that in][only the
minmod scheme was assessed.]{nfnr} Fig.~\ref{fig:PTA} shows the
evolution of the error in the quantity $ r\, D^* / B^{*\,\theta} $ at the
equatorial plane (which is a quantity that should not change with time
with respect to a Lagrangian coordinate system) during the spherical
collapse of a 4/3-polytrope for different $ \{r, \theta\} $ grid
resolutions ($ 80 \times 10 $, $ 160 \times 20 $, and
$ 320 \times 40 $), equally-spaced in the angular direction and
logarithmically spaced in the radial direction. Table~\ref{tab:TT}
gives again numbers for the error and the order of convergence of the
various schemes computed at the end of the simulation
($ t = 20 \mathrm{\ ms} $). In all cases the errors are below 1\%,
even for the coarsest grid, and the order of convergence is higher
than 1 (the presence of local extrema in the radial profiles of some
hydrodynamical variables explains the reduction of the theoretical order
as a built-in feature of total-variation diminishing numerical schemes).
Comparisons between the HLL approximate Riemann solver and the
KT symmetric scheme yield almost identical results \citep[in agreement
with][]{lucas04, shibata_font05, anton06}.


\subsection{Strong spherical explosion}

\begin{figure}[t!]
  \centering
  \resizebox{0.49\textwidth}{!}{\includegraphics*{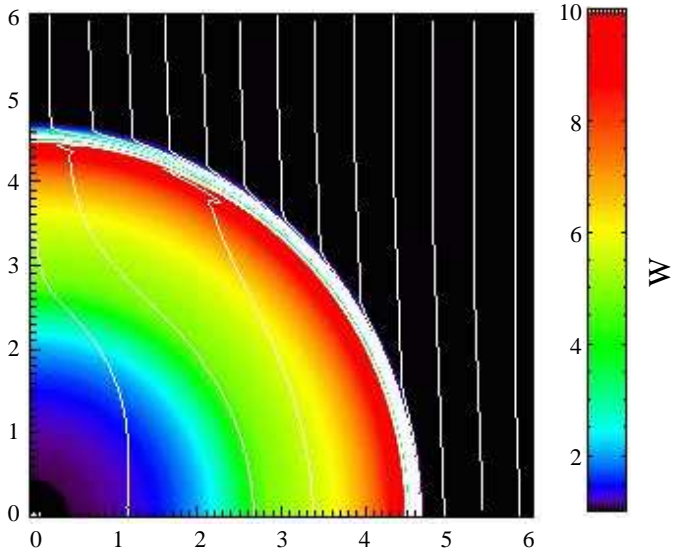}}
  \caption{Spherical explosion test at $ t = 4 $. The Lorentz factor
    $ W $ is color coded, and magnetic field lines are overplotted.}
  \label{fig:SEcolor}
\end{figure}

\begin{figure}[t!]
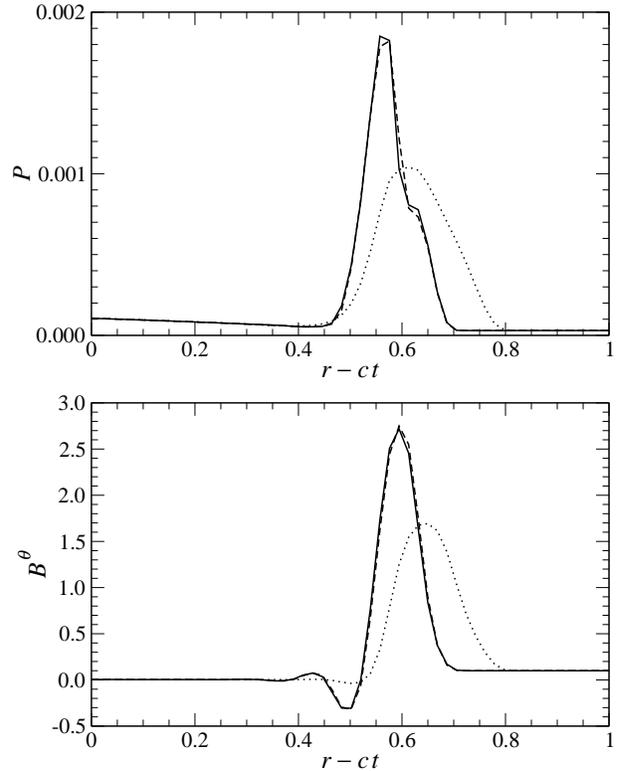

  \centering
  \resizebox{0.46\textwidth}{!}{\includegraphics*{7432f14a.eps}} \\ [0.5 em]
  \resizebox{0.46\textwidth}{!}{\includegraphics*{7432f14b.eps}}
  \caption{Results for the spherical test explosion at $ t = 4 $.
    The plots show profiles for the fluid pressure $ P $ (top panel)
    and the magnetic field component $ B^\theta $ (bottom panel)
    along the equatorial plane using different reconstruction
    schemes: minmod (dotted line), MC (dashed line), and PHM (solid
    line). The grid resolution is $ 320 \times 40 $.}
  \label{fig:SE}
\end{figure}

Explosions are among the most demanding tests for multi-dimensional
codes as they show the ability of numerical schemes to handle shocks.
Since the majority of existing MHD codes are written in Cartesian
coordinates, the most common test is the cylindrical explosion. For
relativistic MHD codes the setup of \citet{komissarov99} for this test
has been used by other authors \citep{delzanna03, leismann05} to
compare different codes. However, in spherical coordinates it is not
possible to impose the symmetries needed for this test. The most
natural choice is thus the spherical explosion. \citet{koessl90}
performed this test in the case of Newtonian MHD. To our knowledge, no
spherical explosions tests have been performed in relativistic
MHD. Therefore, we have designed such a spherical explosion test in
which the initial jump conditions in the variables are the same as for
the test by \citet{komissarov99}. In this way a relativistic shock is
formed which does not occur in the Newtonian case of \citet{koessl90}.

Our test setup consists of an initial explosion zone with $ P = 1.0 $
and $ \rho = 10^{-2} $ for $ r < 0.8 $, surrounded by an ambient gas
with $ P = 3 \times 10^{-5} $ and $ \rho = 10^{-4} $. The explosion region joins
the ambient medium by matching an exponential decline in
a transition region region $ 0.8 < r < 1.0 $. The velocities are
initially zero, and the magnetic field is homogeneous and parallel to
the symmetry axis. The background spacetime is considered to be flat. The
inital data are evolved using an ideal gas EOS with adiabatic index
$ \gamma = 4 / 3 $. We use an evenly spaced grid with a
maximum radius of $ r = 6.0 $. We perform the test for three
resolutions ($ 80 \times 10 $, $ 160 \times 20 $, and
$ 320 \times 40 $) for all reconstruction schemes.

Fig.~\ref{fig:SEcolor} shows the Lorentz factor $ W $ at $ t = 4.0 $.
A strong spherical shock has formed, propagating close to the speed of
light, and as a consequence the magnetic field lines are compressed in
the direction perpendicular to the axis. The results for this test are
qualitatively comparable to the weakly magnetized case in
\citet{komissarov99}. Fig.~\ref{fig:SE} shows radial profiles for
$ P $ and $ B^\theta $ along the equatorial plane at the end of the
simulation, using various reconstruction schemes. These plots are
qualitatively similar to those of the cylindrical explossion
\citep[see e.g. Fig.~B.4 in][]{leismann05}. All numerical schemes
exhibit first order convergence with increasing resolution, as is
expected to happen at shocks. The MC and PHM schemes yield very
similar results, while minmod shows slightly lower values.


\section{Estimation of the growth rates of the \boldmath $ \Omega $-dynamo}
\label{app:odynamo}

To compute the characteristic time scales on which the
$ \Omega $-dynamo mechanism amplifies the magnetic field one has to
study how the wind-up proceeds. Let us consider a stationary rotating
configuration with no meridional flows,
$ v^{*\,r} = v^{*\,\theta} = 0 $ and
$ v^{*\,\varphi} = \Omega^* (r, \theta) \, r \sin \theta $,
where $ \Omega^* (r, \theta) $ stands for the rotation law. Under
these conditions and in the passive field approximation, the induction
equation can be integrated analytically. The solution shows that the
poloidal component of the magnetic field remains constant and the
toroidal component grows linearly with time as
\begin{equation}
  B^{*\,\varphi} (t) =
  B^{*\,\varphi} (t = 0) + t \, \varpi \, \vec{B}^* \cdot
  \vec{\hat{\nabla}} \Omega^*.
  \label{eq:bphi_evolution}
\end{equation}
This equation specifies the toroidal magnetic field at any given time,
provided that the poloidal component is constant and the angular
velocity profile is fixed. For a time $ t \gg B^{*\,0}_\varphi /
(\varpi \vec{B}^* \cdot \vec{\hat{\nabla}} \Omega^*) $,  which is
$ \sim 1 \mathrm{\ ms} $ in our simulations, we can use this
expression to compute the magnetic energy
\begin{equation}
  E_{\mathrm{mag} \, \varphi} \approx
  \int \mathrm{d}^3 \mb{x} \frac{{B^*_{\Omega}}^2}{2}
  \left( \varpi |\vec{\hat{\nabla}} \Omega^*| \right)^2
  \, t^2,
\end{equation}
where $ B^*_\Omega $ is the component of $ \vec{B}^* $ parallel to
$ \vec{\hat{\nabla}} \Omega^* $. The rotation profiles of the final
PNS can be approximated in all our models by the rotation
law~(\ref{rotation_law}) \citep{villain04}. In the Newtonian
limit~(\ref{newtonian_rotation_law}), which is good enough for this
estimate, we can compute an upper limit to the magnetic energy
considering the maximum value of $ | \varpi \vec{\hat{\nabla}}
\Omega^* |_\mathrm{max} = \Omega^*_\mathrm{c} / 2 $, which yields
\begin{equation}
  E_{\mathrm{mag} \, \varphi} \le
  E_{\mathrm{mag} \, \Omega} \,
  \frac{{\Omega^*_\mathrm{c}}^2}{4} \, t^2.
\end{equation}
Therefore, an estimate for the upper limit of the amplification of the
magnetic energy parameter is
\begin{equation}
  \beta_\mathrm{mag}\approx\beta_{\varphi} \leq
  \beta_{\Omega} \frac{{\Omega^*_\mathrm{c}}^2}{4} \, t^2 \leq
  \beta_\mathrm{polo} \frac{{\Omega^*_\mathrm{c}}^2}{4} \, t^2 =
  \left( \frac{t}{\tau_{\Omega}} \right)^2,
  \label{eq:windup}
\end{equation}
where we have defined the time scale for amplification of the magnetic
field by the $ \Omega $-dynamo as
\begin{equation}
  \tau_{\Omega} = \frac{2}{\Omega^*_\mathrm{c} \sqrt{\beta_\mathrm{polo}}}.
  \label{eq:tau_omega}
\end{equation}
This gives us the characteristic time scale in which
$ \beta_\mathrm{mag} $ reaches a value of 1; therefore, the saturation
time $ t_\mathrm{sat} $ should be a fraction of this time. As the
$ \Omega $-dynamo operates by transforming rotational energy into
magnetic energy, the maximum energy can be extracted by the magnetic
field is the one that is contained in the differential rotation of
the core. In accordance with numerical simulations using strong
magnetic fields \citep{obergaulinger_06_a} we estimate this amount to
be 10\% of the total rotational energy, i.e.\ $ \beta_\mathrm{mag}
(t_\mathrm{sat}) = 0.1 \, \beta_\mathrm{rot} (t_\mathrm{sat}) $. We
also assume that the evolution of the magnetic energy parameter is
given by Eq.~(\ref{eq:windup}) and that the energy is conserved, i.e.\
$ \beta_\mathrm{rot} (t) = \beta_\mathrm{rot} (t_0) - \beta_\mathrm{mag} (t) $.

\end{document}